%% file: arhrib.tex
\documentstyle[12pt,epsfig,epsf,feynarts]{article}
\def\ga{\mathrel{\raise.3ex\hbox{$>$\kern-.75em\lower1ex\hbox{$\sim$}}}}
\def\la{\mathrel{\raise.3ex\hbox{$<$\kern-.75em\lower1ex\hbox{$\sim$}}}}

\begin{document}
\begin{flushright}
August 2005 \\
\end{flushright}

\vspace*{0.6cm}
\begin{center}
{\large{{\bf Top and Higgs Flavor Changing Neutral Couplings 
 in two Higgs Doublets Model}}}

\vspace{0.5cm}
Abdesslam Arhrib \\
Physics Division, National Center for Theoretical Sciences,\\ 
PO-Box 2-131 Hsinchu, Taiwan 300
and\\
D\'epartement de Math\'ematiques, Facult\'e des Sciences et Techniques\\
B.P 416 Tanger, Morocco.\\
\end{center}

\begin{abstract}
We study various channels of top and Higgs Flavor Changing 
Neutral Couplings (FCNC) in the two Higgs doublet Model with natural 
flavor conservation (2HDM-I and 2HDM-II). We update known results 
about $t\to c \gamma \ ,\ c Z \ ,\  c g$ and comment on $t\to c h^0$.
The decays $t\to c h^0$ as well as $\{h^0,H^0\}\to \bar{t}c$
are sensitive both to the bottom Yukawa coupling as well as to 
the trilinear scalar couplings $h^0H^+H^-$ and $H^0H^+H^-$.
After imposing unitarity constraints as well as vacuum stability
conditions on scalar sector parameters, in 2HDM-II we found that for  
large $\tan\beta\ga 40$ and rather light charged Higgs mass
$M_{H\pm}\la 150$ GeV, the maximum values allowed for $Br(t\to c h^0)$,
${\rm Br}(H^0\to \bar{t}c)$ and  ${\rm Br}(h^0\to \bar{t}c)$
 are: $8\times 10^{-5}$, $10^{-3}$ and $10^{-4}$ 
respectively. For charged Higgs mass
in the range $[200,300]$ GeV, which can accommodate $B\to X_s\gamma$
constraint if one takes into account large theoretical uncertainties,
the branching ratio of  both $H^0\to
\bar{t}c$ and $h^0\to \bar{t}c$ can still be slightly 
larger than $10^{-5}$.
For $A^0\to \bar{t}c$, its branching ratio is smaller than $\approx
10^{-7}$ in both 2HDM-I and 2HDM-II.
We study also the top-charm associated production 
at $e^+e^-$ colliders and its $\gamma\gamma$ 
option as well as at muon colliders.
It is found that the cross section of $\gamma\gamma\to \bar{t}c$ 
can be of the order $0.01\to 0.1$ fb near threshold region,
while the cross section of $e^+e^-\to \bar{t}c$ is well below 
$10^{-2}$ fb. The situation is slightly better for muon colliders
where a few fb cross section can be reached 
for large $\tan\beta$  and low center of mass energy $\sqrt{s}\la
500$ GeV.
\end{abstract}

\newpage
\section{Introduction}
\label{sec:1}
One of the goals of the next generation of 
high energy colliders, such as the large hadron
colliders LHC \cite{LHC} or the International Linear 
Colliders ILC \cite{TESLA} or muon colliders \cite{Muon}, 
is to probe top flavor-changing neutral 
couplings `top FCNC' as well as the Higgs 
flavor-changing neutral couplings `Higgs FCNC'. 
FCNC of heavy quarks have been 
intensively studied both from the theoretical and experimental point of view. 
Such processes are being well established in Standard Model (SM)
and are excellent probes for the presence 
of new physics effects such as Supersymmetry, 
extended Higgs sector and extra fermions families.
 
Within the SM, with one Higgs doublet, the FCNC $Z t c$ vanish at tree-level
by the GIM mechanism, and the $\gamma \bar{t}c$ and 
$g\bar{t}c$ couplings are zero as a
consequence of  the unbroken $SU(3)_c \times {U(1)}_{\mbox{em}}$ 
gauge symmetry. The Higgs FCNC $H\bar{t}c$ and $Hbs$ couplings also vanish 
due to the existence of only one Higgs doublet.
Both top FCNC and Higgs FCNC are generated at one loop level
by charged current exchange, but they are very suppressed by 
the GIM mechanism.
The calculation of the branching ratios for top decays
yields the SM predictions \cite{sm1}, \cite{sm2}:
\begin{eqnarray}
& & \mathrm{Br}(t \to Zc) = 1.3 \times 10^{-13},
\mathrm{Br}(t \to \gamma c) = 4.3 \times 10^{-13},
\mathrm{Br}(t \to gc) =  3.8 \times 10^{-11},\label{eq1}\\
& & \mathrm{Br}(t \to Hc) = 5.6 \to 3.2 \times 10^{-14}\qquad   
\mathrm{for}\qquad M_H=115 \to 130 \mathrm{GeV}.\label{eq2}\\
& & \mathrm{Br}(H \to \bar{t}c) \approx 1.5\times 10^{-16} \ \
(\mathrm{resp}\ 3\times 10^{-13})\ \ M_H=200\ \  (\mathrm{resp}\ 500) \ 
\mathrm{GeV}\label{eq3}
\end{eqnarray}

In the last decade, there has been intensive activities to 
explore the top FCNC both from the experimental and theoretical point of
view. Experimentally, CDF, D0 \cite{cdfd0} and LEPII \cite{lep2}
collaborations have reported interesting bounds on top FCNC.
These bounds are very weak but will improve in the next years, 
first with Tevatron Run II, and later with the future experiments
such as LHC and/or ILC.\\
From the theoretical side, many SM extensions predict that 
these  top and Higgs FCNC can
be orders of magnitude larger than their SM values (see \cite{rev} for
an overview).  Among those extensions, we cite:\\
i) Models with exotic quarks \cite{exof1}, \cite{exof2}
where one can have branching ratios of the order 
$10^{-2}$, $10^{-5}$ and $5\times 10^{-4}$ respectively for 
$t \to cZ$, $t \to c \gamma $ and $t \to c g$.\\ 
ii) Two Higgs doublet models with and without
Natural Flavor Conservation (2HDM-I, 2HDM-II and 2HDM-III)
 \cite{sm1},\cite{2hdm1},\cite{2hdm2},\cite{2hdm22}, \cite{2hdm3},
\cite{2hdm4}, \cite{abdes}
where the branching ratio can be of the order 
$10^{-6}$, $10^{-7}$, $10^{-5}$ and $10^{-5}$ respectively for 
$t \to cZ$, $t \to c \gamma $, $t \to c g$ and 
$\{h^0, H^0\}\to \bar{b}s$. Note that under some assumptions
on the textures of Lepton Yukawa couplings, the 2HDM can also leads to 
Lepton Falvor violation in Higgs boson decays see Ref.~\cite{sinha} 
for details.\\
iii) and a variety of supersymmetric models 
\cite{susy1}, \cite{susy2}, \cite{susy3}, \cite{maria}. 
In such cases, substantial enhancement can be found
 in  R parity violating MSSM models 
with branching ratios of about
$10^{-3}$, $10^{-5}$ and $10^{-4}$ respectively for 
$t \to c g$, $t \to c \gamma $ and $t \to c Z$.
The corresponding values in the MSSM with R parity conservation 
are of the order $10^{-6}$. 
A particular channel is the top FCNC
coupling $t\to c \Phi$, $\Phi=h^0, H^0, A^0$.
In the case of flavor violation induced by 
gluino, one can reach $10^{-4}$ branching ratio for 
$t\to c h^0$ \cite{susy3}.
 
Hence, top and Higgs FCNC offer a good place to search for new
physics, which may manifest itself if those couplings are observed in
future experiments such as LHC or ILC \cite{LHC,TESLA}. 
At LHC with low luminosity $10 {\rm fb}^{-1}$, 8 million $t\bar{t}$ pairs 
per experiment per year will be produced. This number will increase
by one order of magnitude with the high luminosity option.
Therefore, the properties of top quarks can be examined 
with significant precision at LHC. For top FCNC,
it is possible to reach the following limits \cite{LHC,aguila}:
\begin{eqnarray}
& & {\rm Br}(t\to c H) \leq 4.5\times 10^{-5} \ \ , \ \ 
{\rm Br}(t\to c \gamma ) \leq 3.7\times 10^{-6} \nonumber\\
& & {\rm Br}(t\to c Z ) \leq 7.1\times 10^{-5} \ \ , \ \
{\rm Br}(t\to c g) \leq 10^{-5}
\end{eqnarray}
At ILC the sensitivity is less \cite{aguila}::
\begin{eqnarray}
{\rm Br}(t\to c H) \leq 4.5\times 10^{-5} \ \ , \ \ 
{\rm Br}(t\to c \gamma ) \leq 7.7\times 10^{-6} 
\end{eqnarray}
Consequently, models which can enhance these FCNC couplings close to
the above limits are welcome. 

The aim of this paper is to study top and Higgs FCNC 
couplings in the framework of 
the tree level Flavor Conserving two Higgs doublet 
Models type I and II. We update previous results for  $t \to c\gamma$, 
$t\to  c Z$ and $t\to c g$ \cite{sm1} and revisit
the decay $t \to c h^0$ in the light of 
full unitarity constraints and vacuum stability conditions 
on the Higgs sector.
We also present our results for Higgs FCNC
such as $h^0 \to \bar{t}c$, $H^0 \to \bar{t}c$ and 
$A^0 \to \bar{t}c$ in 2HDM imposing unitarity constraints. 
Moreover, we address the signature of those 
top and Higgs FCNC couplings at $e^+e^-$ Colliders, 
its $\gamma\gamma$ option as well as at future muon colliders.
The isolated top quark signature
may facilitate the search for Higgs FCNC $\Phi \to \bar{t}c$
events and also associate top-charm production.

The paper is organized as follows. 
In the next section, the 2HDM is introduced. Relevant couplings are
given, and  theoretical and experimental constraints 
on 2HDM parameters are discussed.
In the third section, we will study the effect 
of 2HDM on top FCNC decays such as 
$t \to c \gamma$, $t \to c Z$, $t \to c g$ and 
$t \to c h^0$. Section four is devoted to Higgs FCNC couplings:
$Br(\Phi\to \bar{t}c)$, $\Phi=h^0, H^0, A^0$, which are evaluated 
in 2HDM-I and 2HDM-II.
In section 5 we investigate the top-charm associate production 
at $e^+e^-$, its $\gamma\gamma$ options and muon colliders. 
Our conclusions are given in section 6. 

\section{The 2HDM}
Two Higgs Doublet Model (2HDM), is formed by adding an extra complex
$SU(2)_L\otimes U(1)_Y$ scalar doublet to the SM Lagrangian. 
Motivations for such a structure include CP--violation in the Higgs 
sector \cite{weinb1,weinb2} and the fact that some models of
dynamical electroweak symmetry breaking 
yields the 2HDM as their low-energy effective theory \cite{dewsb}.
In particular, the Higgs sector of the Minimal 
Supersymmetric Standard Model (MSSM) takes the form
of a constrained 2HDM.

\begin{figure}[t!]
\begin{center}
\vspace{-1.cm}
\input{hsbb.tex}
\vspace{-8.7cm}
\caption{Generic contribution to $\Phi \to f_1 f_2$ in SM $d_1\to d_{10}$,
in 2HDM $d_{11}\to d_{18}$}
\label{hsbb}
\end{center}
\end{figure}
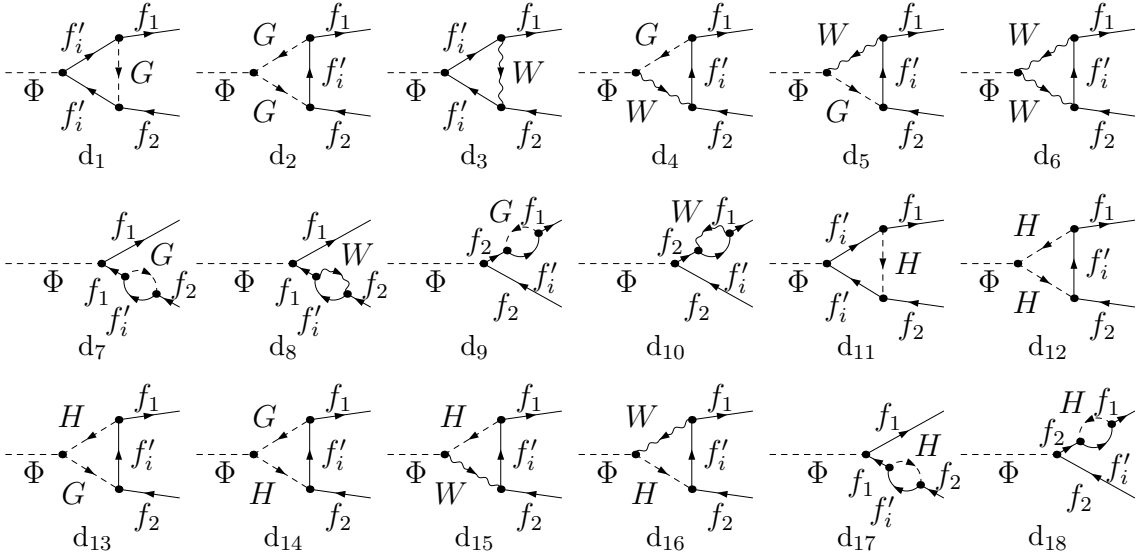

The most general 2HDM scalar potential which is both 
$SU(2)_L\otimes U(1)_Y$ and CP invariant is given by\footnote{It
  exist several ways of writing such a potential see 
 \cite{barroso} for a discussion} \cite{Gun}:
\begin{eqnarray}
 V(\Phi_{1}, \Phi_{2})& & =  \lambda_{1} ( |\Phi_{1}|^2-v_{1}^2)^2
+\lambda_{2} (|\Phi_{2}|^2-v_{2}^2)^2+
\lambda_{3}((|\Phi_{1}|^2-v_{1}^2)+(|\Phi_{2}|^2-v_{2}^2))^2 
+\nonumber\\ [0.2cm]
&  & \lambda_{4}(|\Phi_{1}|^2 |\Phi_{2}|^2 - |\Phi_{1}^+\Phi_{2}|^2  )+
\lambda_{5} [\Re e(\Phi^+_{1}\Phi_{2})
-v_{1}v_{2}]^2+ \lambda_{6} [\Im m(\Phi^+_{1}\Phi_{2})]^2 
\label{higgspot}
\end{eqnarray}
where $\Phi_1$ and $\Phi_2$ have weak hypercharge Y=1, $v_1$ and
$v_2$ are respectively the vacuum
expectation values of $\Phi_1$ and $\Phi_2$ and the $\lambda_i$
are real--valued parameters. 
Note that this potential violates the discrete symmetry
$\Phi_i\to -\Phi_i$ only softly  by the dimension two term
$\lambda_5 \Re e(\Phi^+_{1}\Phi_{2})$. The hard breaking terms 
(dimension four) of the discrete symmetry have been set to zero.\\
The above scalar potential has 8 independent parameters
$(\lambda_i)_{i=1,...,6}$, $v_1$ and $v_2$.
After electroweak symmetry breaking, the combination $v_1^2 + v_2^2$ 
is thus fixed by the electroweak 
scale through $v_1^2 + v_2^2=(2\sqrt{2} G_F)^{-1}$.
We are left then with 7 independent parameters.
Meanwhile,  three of the eight degrees of freedom 
of the two Higgs doublets correspond to 
the 3 Goldstone bosons ($G^\pm$, $G^0$) and  
the remaining five become physical Higgs bosons: 
$H^0$, $h^0$ (CP--even), $A^0$ (CP--odd)
and $H^\pm$. Their masses are obtained as usual
by diagonalizing the mass matrix. 
The presence of charged Higgs bosons will give new contributions
to the one--loop induced top and Higgs FCNC couplings 
(see Fig.~\ref{hsbb}  $d_{11}\to d_{18}$).

It is possible to write the quartic coupling $\lambda_i$ in terms of 
physical scalar masses, mixing angles $\tan\beta$, $\alpha$ and $\lambda_5$ 
as follow \cite{AA}:
\begin{eqnarray}
& & \lambda_4=\frac{g^2}{2 m^2_W} M_{H^\pm}^2 \ \ ,  \ \
\lambda_6=\frac{g^2}{2 m^2_W} M_{A}^2  
\ \ , \ \  \lambda_3=\frac{g^2}{8 m^2_W}
\frac{\sin\alpha \cos\alpha}{ \sin\beta \cos\beta } 
(M_H^2-M_h^2)\ -
\  \frac{\lambda_5}{4} \\
& & \lambda_1 = \frac{g^2}{8 \cos\beta^2 m^2_W} 
[ \cos^2\alpha m^2_H+
\sin^2\alpha m^2_h -
\frac{\sin\alpha \cos\alpha}{\tan\beta}(m^2_H - m^2_h)] 
 -\frac{\lambda_5}{4}(-1 + \tan^2\beta) \nonumber \\ & &
\lambda_2 = \frac{g^2}{8 \sin^2\beta m^2_W} [ \sin^2\alpha m^2_H+
\cos^2\alpha m^2_h -
\sin\alpha \cos\alpha \tan\beta(m^2_H - m^2_h)] 
 +\frac{\lambda_5}{4}(1 - \frac{1}{\tan^2\beta} ) \nonumber
\label{lambda2}
\end{eqnarray}
We are free to take as 7 independent parameters 
$(\lambda_i)_{i=1,\ldots , 6}$ and $\tan\beta$
or equivalently the four physical scalar masses, $\tan\beta$, $\alpha$
and one of the $\lambda_i$. In what 
follows we will take as free parameters:
\begin{eqnarray}
\lambda_5 \quad , \quad M_{h^0} \quad , \quad M_{H^0} \quad , \quad
M_{A^0} \quad , \quad M_{H\pm} \quad , \quad \tan\beta \quad {\rm and} 
\quad \sin\alpha
\label{papa}
\end{eqnarray}

We list hereafter the Feynman rules in the general 2HDM 
for the trilinear scalar couplings relevant for our study. 
They are written in terms of 
the physical masses, $\alpha$, $\beta$ and the soft 
breaking term $\lambda_5$ \cite{AA}:
\begin{eqnarray}
 {H^0H^+H^-}= & &
\frac{-ig}{M_W \sin 2\beta } (
M_{H^0}^2 (\cos\beta^3 \sin\alpha +\sin\beta^3 \cos\alpha)+
M_{H^{\pm}}^2\sin{2\beta} \cos({\beta-\alpha})\nonumber \\ & & -
\sin({\beta+\alpha})\lambda_5 v^2 )\nonumber\label{scalar1}  \\
{H^0H^+G^-}  =& &  \frac{ig}{2 M_W} \sin({\beta-\alpha}) 
(M_{H^0}^2-M_{H^{\pm}}^2)\nonumber\label{scalar2} \\
{h^0H^+H^-}  = && \frac{- ig}{M_W \sin 2\beta} ( 
 M_{h^0}^2(\cos{\alpha}\cos{\beta}^3-
\sin{\alpha}\sin{\beta}^3)
 +M_{H^{\pm}}^2\sin{2\beta}\sin({\beta-\alpha})\nonumber
\label{scalar33} \\ && -
\cos({\beta+\alpha}){\lambda_5} v^2)\label{scalar3}   \\
{h^0H^+G^-} =&&\frac{-ig}{2 M_W}  \cos({\beta-\alpha}) 
(M_{h^0}^2-M_{H^{\pm}}^2) \nonumber\label{scalar4}  \\
 {A^0H^+G^-}  = & & \frac{-g}{2 M_W}  
(M_{H^{\pm}}^2 - M_A^2)\qquad , \ \ v^2 = \frac{2M_W^2}{g^2} \label{scalar5}
\end{eqnarray}
We need also the couplings of the scalar bosons to a pair of fermions
both in 2HDM-I and 2HDM-II. In these couplings, the relevant terms 
are as follows:
\begin{eqnarray}
& & h^0\bar{t}t \propto M_t\frac{\cos\alpha}{\sin\beta} \ \ \ , \ \ \ 
H^0\bar{t}t \propto M_t \frac{\sin\alpha}{\sin\beta}\ \ \ , \ \ \ 
A^0\bar{t}t \propto \frac{M_t}{\tan\beta} \ \ {\rm 2HDM-I , II}
\nonumber\label{coupl1}\\
& & h^0\bar{b}b \propto M_b\frac{\cos\alpha}{\sin\beta} \ \ \ , \ \ \ 
H^0\bar{b}b \propto M_b \frac{\sin\alpha}{\sin\beta}\ \ \ , \ \ \ 
A^0\bar{b}b \propto \frac{M_b}{\tan\beta}\ \ {\rm 2HDM-I }
\nonumber\label{coupl2}\\
& & h^0\bar{b}b \propto M_b\frac{\sin\alpha}{\cos\beta} \ \ \ , \ \ \ 
H^0\bar{b}b \propto M_b \frac{\cos\alpha}{\cos\beta}\ \ \ , \ \ \ 
A^0\bar{b}b \propto M_b\tan\beta \ \ {\rm 2HDM-II }\nonumber\label{coupl3}\\
& & (H^-\bar{b}t)_L \propto \frac{M_b}{\tan\beta} \ \ \ , \ \ \ 
(H^-\bar{b}t)_R \propto \frac{M_t}{\tan\beta}  \ \ {\rm 2HDM-I } 
\nonumber\label{coupl4}\\
& & (H^-\bar{b}t)_L \propto {M_b}{\tan\beta} \ \ \ , \ \ \ 
(H^-\bar{b}t)_R \propto \frac{M_t}{\tan\beta}  \ \ {\rm 2HDM-II } 
\label{coupl5}
\end{eqnarray}

To constrain the scalar sector parameters we will use both 
vacuum stability conditions as well as
tree level unitarity constraints. In our study, we use the vacuum 
stability conditions from \cite{vac1,vac2}, which are given by:
\begin{eqnarray}
\nonumber
& \lambda_1 + \lambda_3 > 0\;,\quad\quad \lambda_2 + \lambda_3 > 0\;, 
\nonumber\\
& 2\sqrt{(\lambda_1 + \lambda_3)(\lambda_2 + \lambda_3)} 
+2\lambda_{3} +\lambda_4 + {\rm{min}}
\left( 0, \lambda_{5} - \lambda_{4} , \lambda_{6}-\lambda_{4}
 \right) >0  \label{vac}
\end{eqnarray} 

It is well known that the unitarity bounds coming from a tree-level 
analysis put severe constraints on all scalar trilinear and 
quartic couplings~\cite{LTQ,kan,abdesunit}. 
The tree level unitarity bounds are derived with 
the help of the equivalence theorem, which itself is a 
high-energy approximation where it is assumed that the 
energy scale is much larger than the $Z^0$ and $W^\pm$ 
gauge-boson masses.  To derive these unitarity constraints,
we consider the following neutral scattering processes
\begin{eqnarray}
&& W_L^+ W_L^-, Z_L Z_L, W_L^+ H^-, W_L^- H^+, Z_L h^0, Z_L H^0, Z_L A^0, 
A^0 A^0, H^+ H^-,  A^0 h^0, A^0 H^0,\nonumber\\ && h^0 h^0, h^0 H^0,
H^0 H^0,
\end{eqnarray}
as well as the charged channels:
\begin{eqnarray}
W_L^+ h^0, W_L^+ H^0, W_L^+ A^0, W_L^+ Z_L, Z_L H^+, H^+ h^0, 
H^+ H^0, H^+ A^0
\end{eqnarray}
It has been demonstrated, first in 2HDM with exact
 discrete symmetry \cite{kan}  and later in 2HDM with softly broken 
discrete symmetry  \cite{abdesunit},
that the above neutral scattering processes lead to the 
following 14 unitarity constraints\footnote{We use the rather conservative
unitarity bound $16\pi$ instead of $8\pi$}:
\begin{eqnarray}
|a_{\pm}| \ , \ |b_{\pm}| \ , \ |c_{\pm}| \ , \ |d_{\pm}| \ , \ |f_{\pm}| \ , 
\ |e_{1,2}| \ , \ |f_{1,2}| \  \ \leq 16\pi
\label{constraint}
\end{eqnarray}
with
\begin{eqnarray}
& & a_{\pm} =3 (\lambda_1 + \lambda_2 + 2 \lambda_3) \pm
\sqrt{9 (\lambda_1 - \lambda_2)^2 +
      (4 \lambda_3 +  \lambda_4 + 0.5(\lambda_5 + \lambda_6))^2}\nonumber\\
& & b_{\pm}=\lambda_1 + \lambda_2 + 
2 \lambda_3 \pm \sqrt{ (\lambda_1 - \lambda_2)^2 +
0.25(-2 \lambda_4 + \lambda_5 + \lambda_6)^2}\nonumber\\
& & c_{\pm}=\lambda_1 + \lambda_2 + 
2 \lambda_3 \pm \sqrt{(\lambda_1 - 
\lambda_2)^2 + 0.25(\lambda_5 - \lambda_6)^2} \nonumber\\
& & e_1=2 \lambda_3 - \lambda_4 - \frac{\lambda_5}{2} + \frac{5}{2}
  \lambda_6 \quad , \quad e_2=2 \lambda_3 + \lambda_4 - 
\frac{\lambda_5}{2} + \frac{1}{2}\lambda_6 \nonumber\nonumber\\
& & f_+=2 \lambda_3 - \lambda_4 +  \frac{5}{2}\lambda_5 - \frac{1}{2}
  \lambda_6\quad , \quad f_-=2 \lambda_3 + \lambda_4 + 
\frac{1}{2}\lambda_5 - \frac{1}{2}\lambda_6\nonumber\\
& & f_1=f_2=2 \lambda_3 + \frac{1}{2} \lambda_5 + \frac{1}{2}
  \lambda_6
\label{pert}
\end{eqnarray}
In addition, it has been noticed in \cite{abdesunit}, that 
the inclusion of charged scattering processes leads to one extra
unitarity constraint and is given by (see
\cite{abdesunit} for more details):
\begin{eqnarray}
p_1 =2 (\lambda_3 +  \lambda_4 ) - 0.5(\lambda_5 +\lambda_6)
\leq 16\pi
\end{eqnarray}
This additional constraint together with the others play
an important role in constraining the parameter space of the 2HDM
\cite{abdesunit}.
In our analysis, we take into account also the following 
constraints when the independent parameters are varied.
From the theoretical point of view:\\
$i)$ The extra contributions to the $\delta\rho$ parameter from the Higgs
scalars should not exceed the current limits from precision 
measurements \cite{PDG}: $|\delta\rho|\la 0.001$.
Such extra contribution to $\delta\rho$
vanishes in the limit $M_{H^\pm}=M_A$ which gives $\lambda_4=\lambda_6$. 
Under this constraint the 2HDM scalar 
potential is $O(4)$ symmetric \cite{negative}. In this case
$(H^+, A^0, H^-)$ form a triplet under the residual global $SU(2)$ of
the Higgs potential. It is this residual symmetry which ensures that 
$\rho$ is equal to unity at tree level. 
It is clear from this discussion,
that $\delta\rho$ will be violated once the splitting 
between $H^\pm$ and $A^0$ is large.\\
$ii)$ From the requirement of perturbativity for the
top and bottom Yukawa couplings, $\tan\beta$ is 
constrained to lie in the range $0.1\leq \tan\beta \leq 70$ \cite{berger}. \\
$iii)$ It has been shown in~\cite{bsg} that 
for models of the type 2HDM-II, data on $B\to X_s \gamma$ 
impose a lower limit of $M_{H^\pm} \ga 350$\,GeV.
In type I 2HDM, there is no such  constraint on the 
charged Higgs mass \cite{bsg}. 
Recently, it has been shown, using a renormalization group improved
calculation of the $B\to X_s\gamma$ branching ratio, that the scale
$\Delta=m_b-E_0$ introduces significant 
perturbative uncertainties \cite{neubert}. Such large theory
uncertainties may have important implications for searches 
of new physics in radiative B decays. It is pointed out in 
\cite{neubert}, that the lower bound on the charged Higgs mass is
reduced to about 200 GeV. 
In our numerical analysis we will ignore these
constraints and allow $M_{H\pm}\la 200$ GeV 
in order to localize regions in parameter
space where the branching ratios are sizeable.\\
$iv)$ The most recent data on the $a_\mu=(g-2)_\mu/2$ indicates 
\cite{gm2}
\begin{eqnarray}
\Delta a_\mu = a_\mu^{\rm exp} - a_\mu^{\rm SM}=(23.9\pm 10)\times
10^{-10}\label{gm2}
\end{eqnarray}
If the SM prediction is based on the hadronic contribution from 
$e^+e^-$, the above result (\ref{gm2}) shows 2.4 $\sigma$ deviation
from SM. The dominant 2HDM contribution to $(g-2)_\mu$ come from one
loop diagram with Higgs exchange \cite{hab,mariaa} and also from
two-loop Barr-Zee diagram with heavy fermion running in 
the upper loop \cite{dar,otto}. In both cases, sizeable contribution
to $(g-2)_\mu$ appear only for relatively light neutral Higgs and
large $\tan\beta$. In our study we will assume that all neutral Higgs
are heavier than 100 GeV (see below). This assumption make the one
loop contribution to $(g-2)_\mu$ within experimental range for any
value of $\tan\beta\la 70$ and the mixing $\alpha$. The 
two-loop Barr-Zee diagram has been studied extensively in \cite{otto}.
It has been shown in \cite{otto}, that for neutral Higgs bosons
heavier than 100 GeV and $\tan\beta$ of the order 60, $(g-2)_\mu$
two-loop Barr-Zee diagram are within the experimental
range. Therefore, since we will assume that all Higgs are heavier than
100 GeV we will ignore $(g-2)_\mu$ constraint in what follows.
\\ $v)$ From the experimental point of view,
the combined null--searches from all four CERN LEP collaborations derive the 
lower limit $M_{H^{\pm}}\ge 78.6$ GeV $(95\%\, CL)$, a limit
which applies to all models in which BR($H^{\pm}\to \tau\nu_{\tau}$)+
BR($H^{\pm}\to cs$)=1. For the neutral Higgs bosons,
OPAL collaboration has put a limit on 
$h^0$ and $A^0$ masses  
of the 2HDM. They conclude that the regions
$1\la M_h \la 44$ GeV and $12\la M_A \la 56$ GeV 
are excluded at 95\% CL independent of $\alpha$ and $\tan\beta$
\cite{opal}. In what follows, we will assume that all Higgs masses are 
greater than 100 GeV.

\section{Top FCNC in 2HDM}
In this section, we will discuss the 2HDM contribution to 
top FCNC couplings $t \to c \gamma$, $t\to c g$ and 
$t\to c Z$ as well as $t \to c h^0$. In the case of $t \to c \gamma $, 
$t\to c g$ and $t\to c Z$ couplings, we reproduce and update the 
results of ref.~\cite{sm1},
while for the case of $t \to c h^0$ couplings, we revisit the study done by 
S. B\'ejar et al \cite{2hdm3} in the light of the full set of tree 
level unitarity  constraints and vacuum stability conditions.\\
The full one-loop calculation presented here is done in the 'tHooft
gauge with the help of 
FormCalc \cite{FA2} and FF packages \cite{FF}.
Before presenting our numerical results, we would like to mention that
we have adopted the following inputs:
\begin{eqnarray}
& & \alpha=1/137.035989 \ , \  M_W=80.45 \  ,  \ M_Z=91.1875\ {\rm GeV} \ ,
\ s_W^2=1-M_W^2/M_Z^2\\
& & M_t=174.3 \  , \  M_b=4.7 \  , \  M_c=1.5 \ , 
\ M_s=0.2 \ {\rm GeV} \ \ , \ \
V_{cb}=0.04 \ \ , \ \ \alpha_s(M_t) \approx 0.105\nonumber
\end{eqnarray}
\begin{figure}[t!]
\smallskip\smallskip 
\vskip-3.8cm
\centerline{{
\epsfxsize3.1 in 
\epsffile{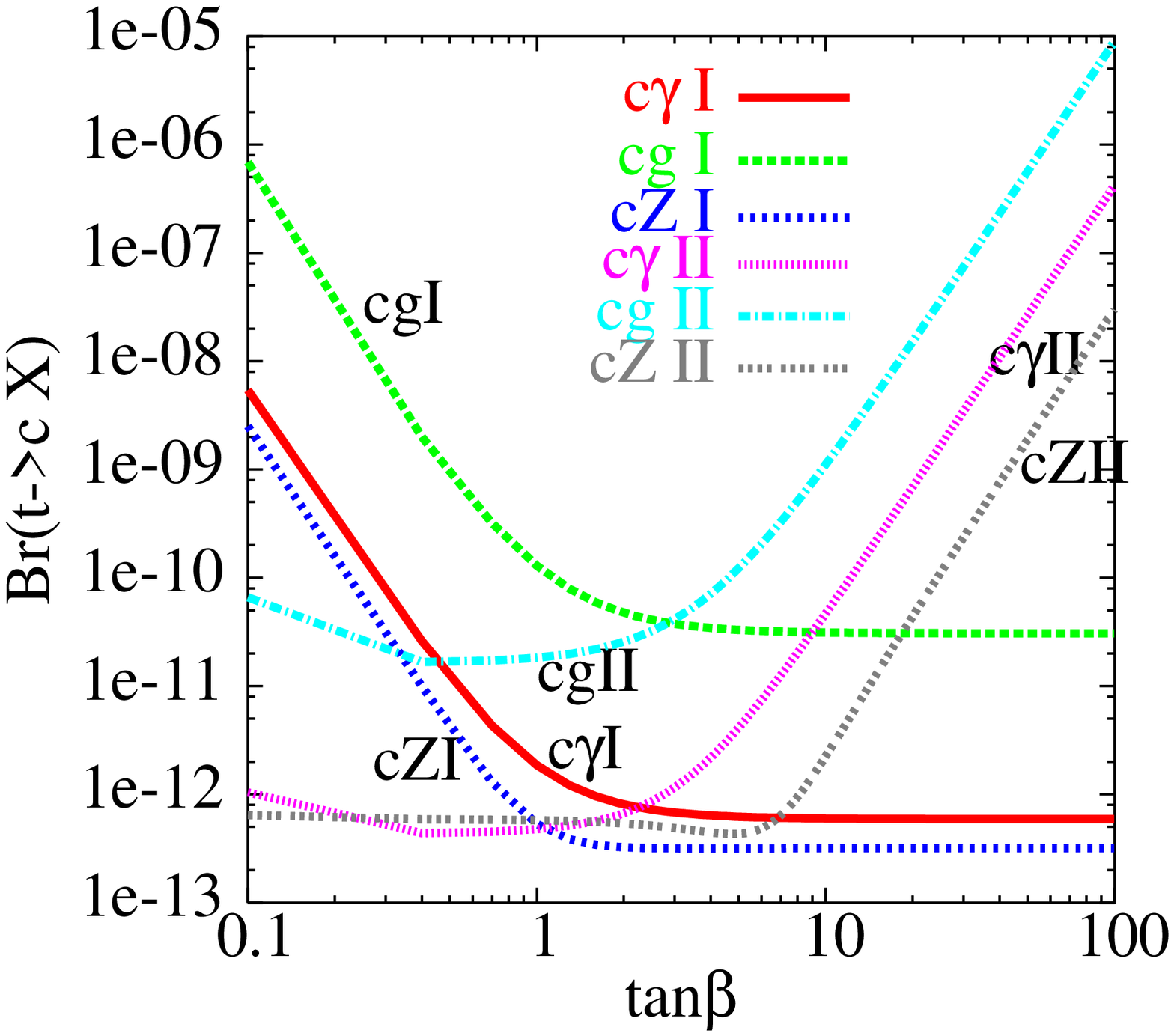}}  \hskip0.4cm
\epsfxsize3.1 in 
\epsffile{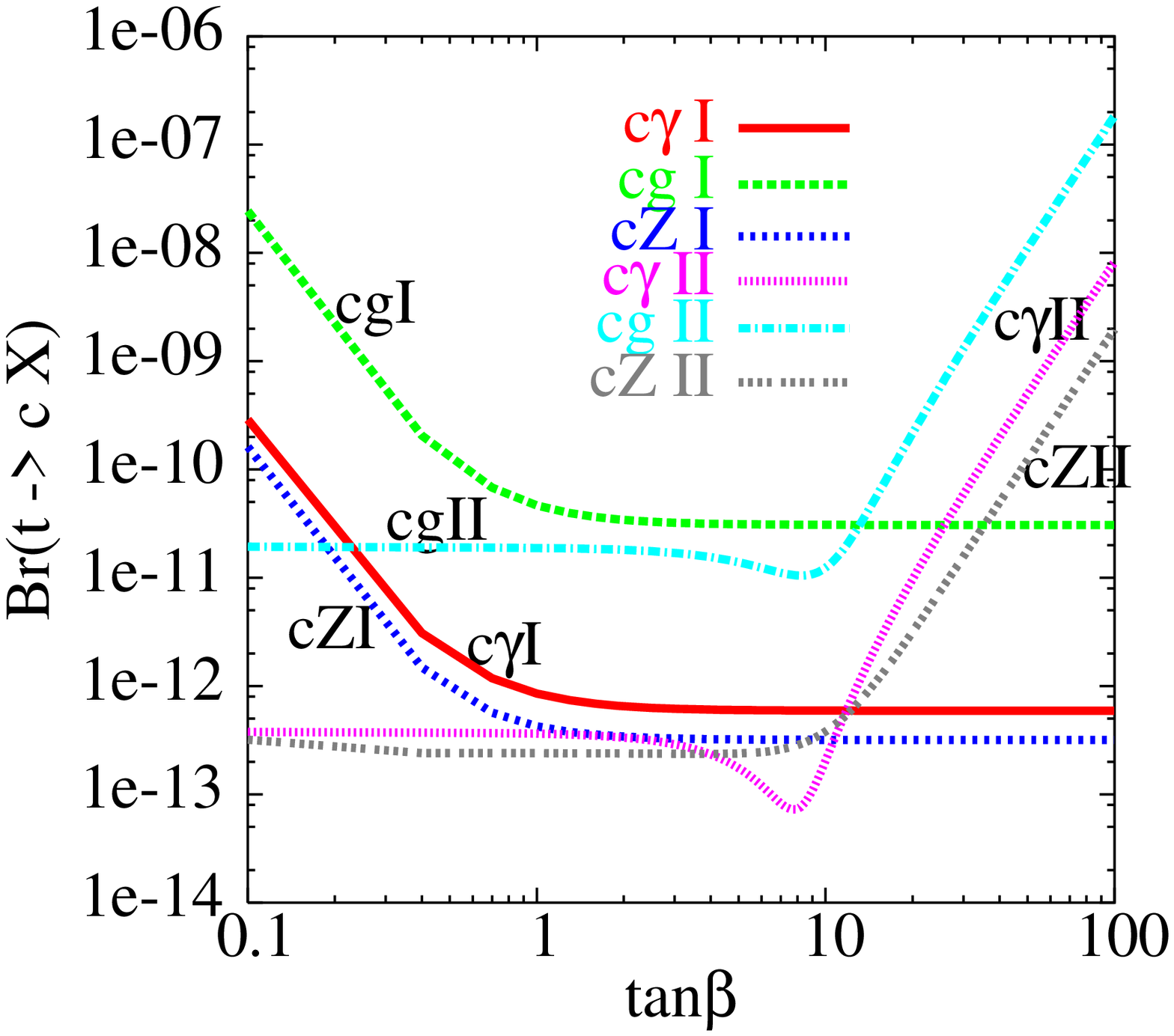} }
\smallskip\smallskip
\caption{$t \to c \gamma $, $t\to c g$ and 
$t\to c Z$ as function of $\tan\beta$ in 2HDM-I and 2HDM-II for 
$M_{H\pm}=120$ GeV (left), $M_{H\pm}=250$ GeV (right)}
\label{fig5}
\end{figure}
In the 2HDM type I and II, the top FCNC couplings $t \to c \gamma $, 
$t\to c g$ and $t\to c Z$ are sensitive only to the charged 
Higgs mass and $\tan\beta$. We update the plots of ref.~\cite{sm1}
for 2 values of charged Higgs mass $M_{H\pm}=120$ and $250$ GeV
and $\tan\beta \in [0.1,100]$. Our results, in perfect 
agreement with \cite{sm1}, are illustrated in Fig.~(\ref{fig5}).
In these plots we show
the Branching ratios of $t \to c \gamma $, 
$t\to c g$ and $t\to c Z$ both in 2HDM-I and 2HDM-II as 
 a function of $\tan\beta$ for $M_{H\pm}=120$ GeV (left plot) and
$M_{H\pm}=250$ GeV (right plot).
As  can be seen from these plots, only at small $\tan\beta$
(for 2HDM-I) and large $\tan\beta$ (for 2HDM-II),
 can the branching ratios be in the range $10^{-8}-10^{-6}$. 
Even $t\to c g$ decay which is proportional to the strong coupling $\alpha_s$
does not exceed a branching ratio of $10^{-5}$ at large $\tan\beta\ga 70$.
Therefore, if one of those top FCNC emerge at LHC,
with a rate larger than $10^{-5}$ that would definitely
attest the existence of new sources of flavor violation other than the one 
induced by the 2HDM structure.

Let us now concentrate on $t\to c h^0$.
This top FCNC coupling depends on the scalar parameters
through the pure scalar couplings: $h^0 H^+ H^-$ and $h^0 G^+ H^-$. 
These scalar couplings eqs.~(\ref{scalar5}),
depend on the seven scalar parameters 
listed above eq.~(\ref{papa}) and may be sources of enhancement. 
Feynman diagrams which depend on these pure scalar couplings
are depicted in  Fig.~(\ref{hsbb}): $d_{12}$, $d_{13}$ and $d_{14}$.
Top FCNC $t\to c h^0$ depends also on the bottom Yukawa coupling
through $H^+\bar{b}t$ and/or $h^0b\bar{b}$ interaction 
which could also enhance the width of 
$t\to c h^0$ for large $\tan\beta$. These kinds of diagrams can be seen
in $d_{1,3}$ and $d_{11\to 18}$.\\
At large $\tan\beta$, it is expected that the main enhancement 
comes from diagrams of the type $d_{11}$ and $d_{12}$. 
If we neglect the charm mass
and right coupling $(H^-\bar{f}t)_R\propto 1/\tan\beta$, 
which are suppressed for large $\tan\beta$, the amplitudes 
$M_{11}$ of $d_{11}$ and $M_{12}$ of $d_{12}$
are given by:
\begin{eqnarray}
&& M_{11}\approx \frac{h^0f\bar{f}}{16\pi^2} 
M_t V_{ft} V_{cf}^*\ \ (H^-\bar{f}t)_L\ \  (H^-\bar{f}c)_L\ \ 
f_1(M_{H\pm}^2,M_t^2,M_f^2,M_h^2) {v}_t\frac{1+\gamma_5}{2}u_c\label{m11}\\
&& M_{12}\approx \frac{h^0H^+H^-}{16\pi^2} 
M_t V_{ft} V_{cf}^*\ \  (H^-\bar{f}t)_L\ \  (H^-\bar{f}c)_L \ \ 
f_2(M_{H\pm}^2,M_t^2,M_f^2,M_h^2) {v}_t\frac{1+\gamma_5}{2}u_c\label{m12}
\end{eqnarray}
$f_1$ and $f_2$ are form factors and $f$ is the internal quark.
The couplings $h^0f\bar{f}$, $h^0H^+H^-$ and $(H^-\bar{f}t)_L$ are
given in eqs.(\ref{scalar5}, \ref{coupl5}). 
Since in 2HDM-II, $h^0f\bar{f}$ behaves like $m_f\tan\beta$ at 
large $\tan\beta$ and the coupling $h^0H^+H^-$ is 
enhanced for large $\tan\beta$ (see
discussion below), one can expect that the amplitude of those diagrams 
grows like $\tan^3\beta$. But care  has to be taken, since the 
$\tan^3\beta$ enhancement of the amplitude may be 
reduced by the CKM elements, such as $V_{cb}$,
as well as by the GIM cancellation.
\begin{figure}[t!]
\smallskip\smallskip 
\vskip-.1cm
\centerline{{
\epsfxsize3.18 in 
\epsffile{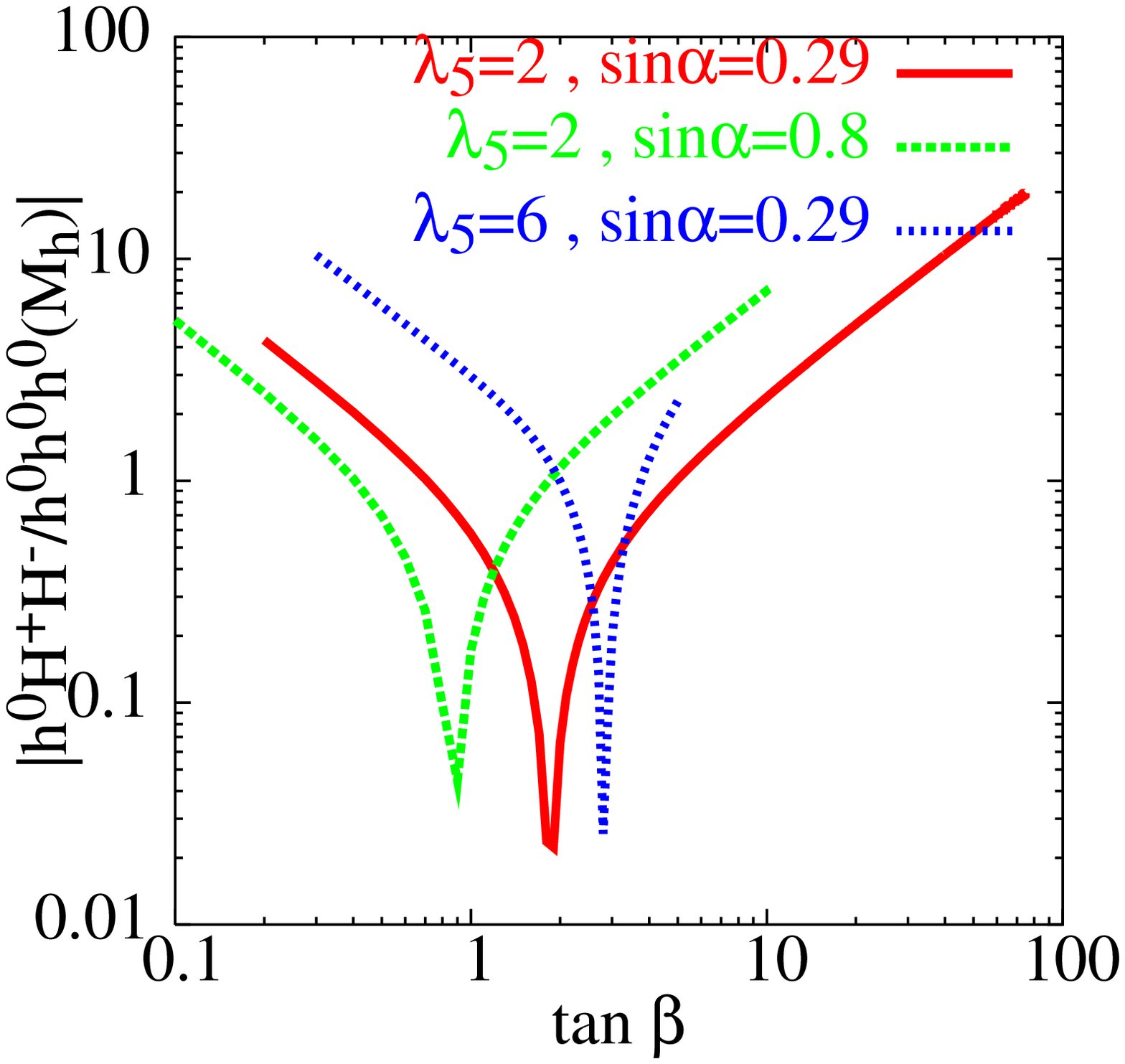}} 
\hskip-.01cm
\epsfxsize3.18 in 
\epsffile{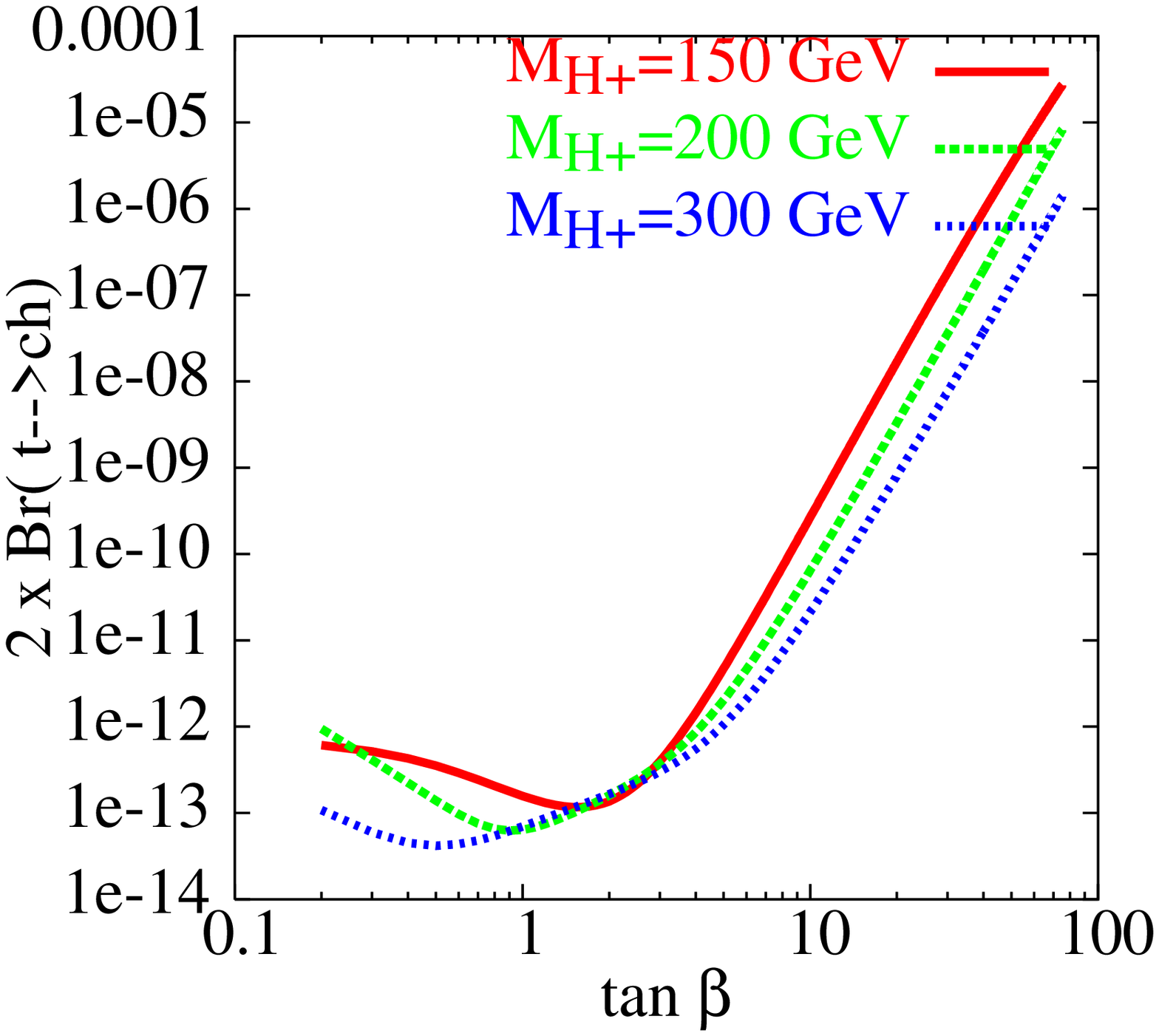}}
\smallskip\smallskip
\caption{
{\bf Left}, Absolute value of $h^0H^+H^-$ 
coupling normalized to SM coupling
$h^0h^0h^0$ taken at $M_h^0=125$ GeV, the other parameters are fixed as
$M_{H\pm}=150$, 
$M_{A^0}=215$ and $M_{H^0}=250$ GeV,
$\lambda_5=2$ . {\bf Right}, $2\times Br(t\to c h^0)$ in 2HDM-II
as function of $\tan\beta$ for several values of $M_{H\pm}$ and
$M_{h^0}=125$, $M_{A^0}=215$, $M_{H^0}=250$ GeV, $\lambda_5=2$ and 
$\sin\alpha=0.29$}
\label{fig6}
\end{figure}
\begin{figure}[t!]
\smallskip\smallskip 
\vskip-.1cm
\centerline{{
\epsfxsize3.18 in 
\epsffile{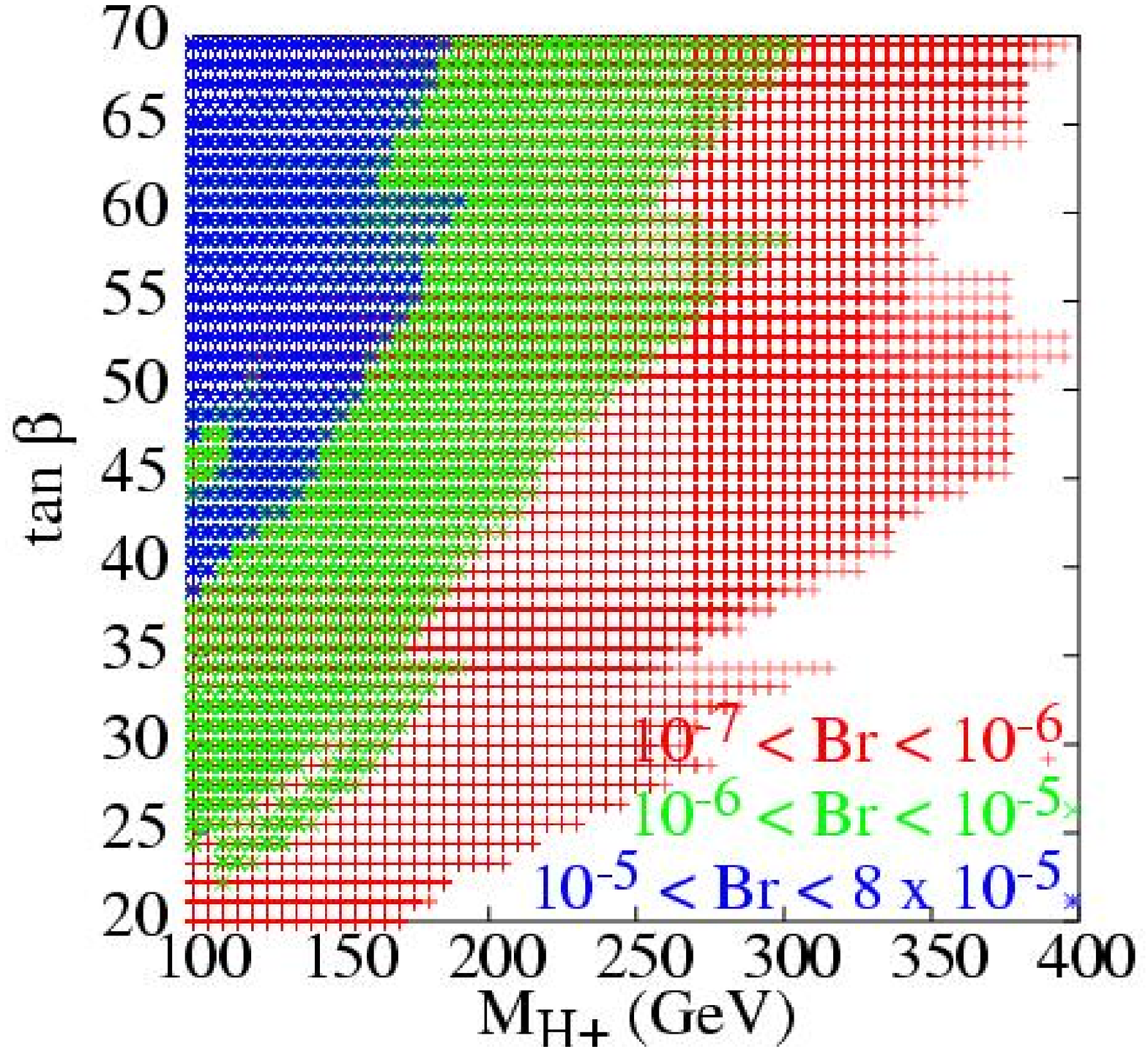}} 
\hskip-.01cm
\epsfxsize3.18 in 
\epsffile{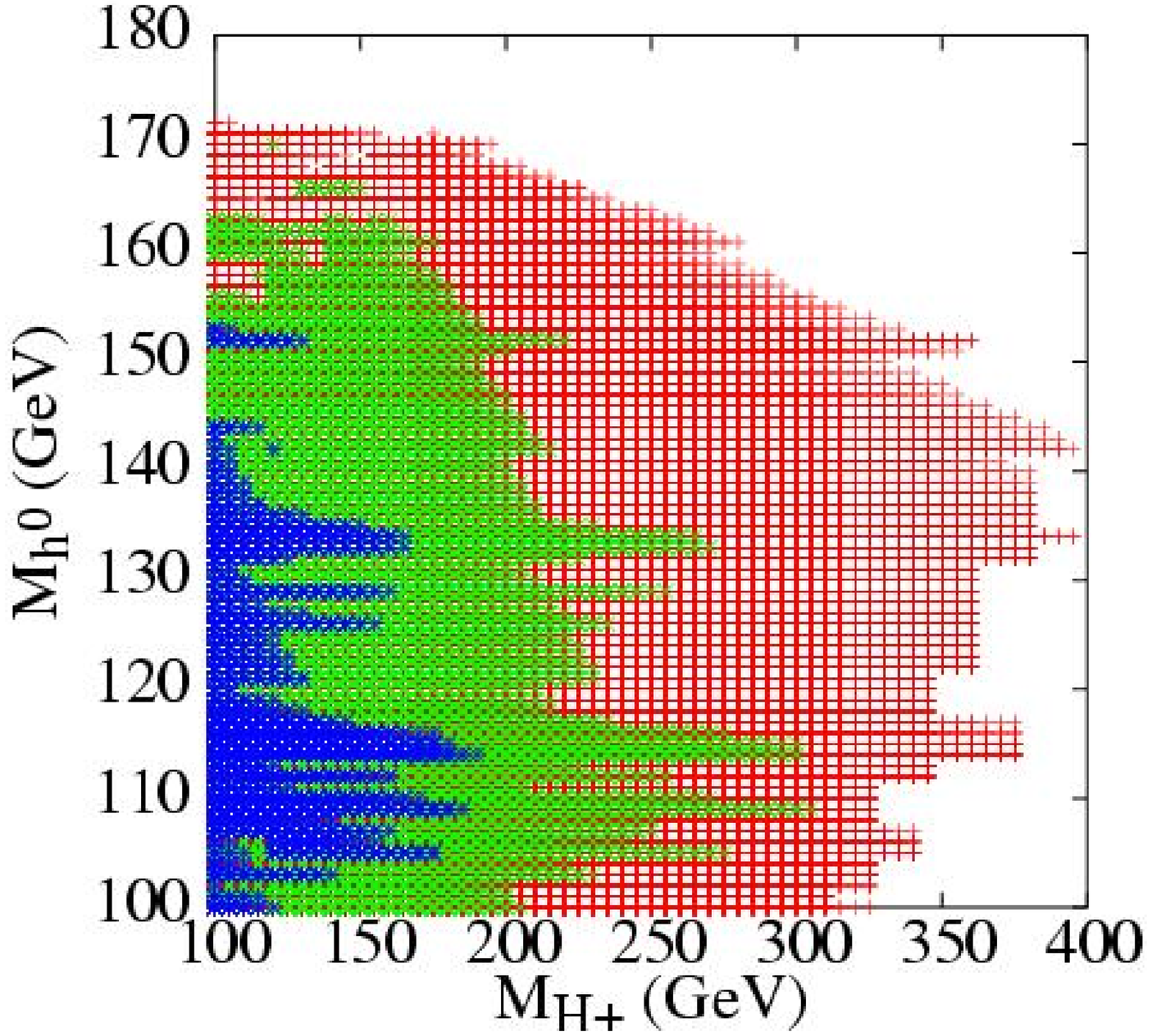}}
\smallskip\smallskip
\caption{Scatter plots for ${\rm Br}(t\to c h^0)$ in the 
  ($\tan\beta$, $M_{H\pm}$) plan (left) and ($M_{h^0}$, $M_{H\pm}$)
  plan. We scan over: $100 \la M_{h^0} \la 175$, 
$100 \la M_{H^0,A^0,H\pm} \la 600$ GeV, $-1\la \sin \alpha \la 1$ and 
$20\la \tan\beta \la 70$ and $\lambda_5=M_A^2/v^2$.}
\label{fig66}
\end{figure}
When computing the branching ratio $Br(t\to c h^0)$, 
in the top width we include both the tree level decay $t\to W^+ b$ 
as well as  $t\to H^+ b$ when it is open.
We make a comparison of our results with the results obtained by 
B\'ejar et al \cite{2hdm3}, and find perfect agreement. However
the parameters set chosen by Ref.~\cite{2hdm3} to illustrate 
their results violate  both unitarity  and perturbativity 
constraints on $\lambda_i$\cite{kan,abdesunit}.

We have performed a systematic scan over the full parameter
in the range allowed by the theoretical constraints such as unitarity
constraints and vacuum stability conditions
as well as by the experimental data.  
As one can see from the scatter
plot Fig.~(\ref{fig66}), 
the branching ratio of  $t\to c h^0$ remains less than $8\times
10^{-5}$ for light charged Higgs mass and large $\tan\beta$.

Let us start by discussing the order of magnitude of the triple
coupling $h^0H^+H^-$. In the left plot of Fig.~(\ref{fig6})
we illustrate the  
ratio $R=|h^0H^+H^-/h^0h^0h^0|$, with  $h^0h^0h^0$ being the SM
coupling given by $h^0h^0h^0=3 g M_{h^0}^2/(2 M_W)$. 
In this plot, we use $M_{h^0}=125 $ GeV and 
the other parameters  take the values
$M_{H\pm}=150$,  $M_{A^0}=215$ and $M_{H^0}=250$ GeV.
As it can be seen from this plot, unitarity constraints allow 
$\tan\beta$  to be as large as 70 only for 
$\sin\alpha\approx 0.29$ and $\lambda_5=2$. 
For large $\sin\alpha=0.8$
unitarity already breaks down for $\tan\beta\approx 10$. 
It is clear that the coupling $h^0H^+H^-$
reach its maximal value for large $\tan\beta$. The minimum 
observed in the $h^0H^+H^-$ coupling 
is due to a cancellation between the different terms which contribute
to $h^0H^+H^-$ (eq.(\ref{scalar3})). 
Such a minimum depends of course 
on the values of $\lambda_5$ and $\sin\alpha$.
Since the coupling $h^0H^+H^-$ depends linearly on $\lambda_5$,
it is obvious that this coupling gets its
largest value for the largest value of $\lambda_5$ tolerated by
perturbativity: $\lambda_5\approx 8 \pi$. 
However, unitarity requires that for large $\lambda_5$, 
large $\tan\beta$ is not allowed 
(see left plot Fig.~\ref{fig6} for $\lambda_5=6$).  

Having the order of magnitude of $h^0H^+H^-$ in mind, 
we show in the right panel of Fig.~(\ref{fig6}) 
the Branching ratio of $t\to c h^0$ as a function of $\tan\beta$ for 
several values of $M_{H\pm}$ with
$M_{h^0}=125$, $M_{A^0}=215$, $M_{H^0}=250$ GeV, 
$\lambda_5=2$ and $\sin\alpha=0.29$. 
For large $\tan\beta$ and light charged Higgs $M_{H\pm}=150$ GeV,
Br($t\to c h^0$) can reach value larger than $10^{-5}$ while for
 $M_{H\pm}=200$ GeV, the Br($t\to c h^0$) is already less than 
$10^{-5}$ even for large $\tan\beta\approx 70$.\\
This enhancement is mainly attributed to the triple 
scalar coupling $h^0H^+H^-$ as well as to 
the bottom Yukawa coupling which are both enhanced  for large 
$\tan\beta$ as one can see from eqs.(\ref{m11},\ref{m12}). \\
We also provide scatter plots Fig.~(\ref{fig66}) for ${\rm Br}(t\to c h^0)$ 
in ($M_{H\pm}$, $\tan\beta$) and  ($M_{H\pm}$, $M_{h^0}$) plans.
There is only a small window for large $\tan\beta$ and $M_{H\pm}\la
180$ GeV where ${\rm Br}(t\to c h^0)$ can be larger than $10^{-5}$.
As argued before, due to the  large theoretical uncertainties
in the $B\to X_s\gamma$ calculation, the lower bound on the 
charged Higgs mass is reduced to about 200 GeV \cite{neubert}.
From the left panel of Fig.~\ref{fig66}, it is illustrated that for charged
Higgs mass $M_{H\pm}\in [200,300]$ GeV and large $\tan\beta$, 
 ${\rm Br}(t\to c h^0)\in [10^{-6},10^{-5}]$.
One concludes that 
the top FCNC $t\to ch^0$ in 2HDM can reach a
detectable rate at LHC with the high luminosity option.

\section{Higgs FCNC in 2HDM}
\subsection{Higgs FCNC in SM}
Before presenting our results in 2HDM, we would like to comment on
 the Branching ratio of $H\to {\bar{t}}c$ in the SM.
Recently, an estimation using dimensional analysis and power counting,
 appeared both for $Br(H\to {\bar{s}}b)$
 \cite{bdgs}  and $Br(H\to {\bar{t}}c)$ \cite{2hdm4} while
in \cite{abdes}, the full diagrammatic calculation has been done.  
Here we comment on results 
for $Br(H\to {\bar{t}}c)$  and $Br(H\to {\bar{b}}s)$\cite{abdes}.
\\
As expected the branching ratios of these decays 
are very suppressed due to GIM mechanism \cite{2hdm4,abdes,bdgs}.
In the case of $H\to {\bar{t}}c$, the internal fermions are bottom,
strange and down quarks with masses close to degeneracy. 
The GIM suppression in this case is very severe, and the Br is of the order
$10^{-13}$.
For $H\to \bar{b}s$ the internal fermions are top,
charm and up quarks and with the large top mass as 
well as  large splitting between
 the internal quarks, the Branching ratio of $H\to \bar{b}s$ is  in
the range $10^{-9}$--$10^{-7}$. $H\to \bar{b}s$ has another advantage 
over $H\to \bar{t}c$ which is the fact that $H\to \bar{b}s$ can be 
open for Higgs mass $M_H<2 M_W$ and hence the decay width of 
the Higgs can be very narrow
which can enhance the Branching ratio for $M_H\la 2 M_W$.
The decay $H^0\to \bar{t}c$ is open only when $M_{H}>m_t+m_c 
\approx 2 M_Z $. 
For a Higgs mass in this range, the decay channels 
$H^0\to \{ W^+W^-, Z^0Z^0\}$ are already open and so the total width 
of the Higgs is no longer narrow.
Such a large decay width may reduce  the Branching ratio of 
$H^0\to \bar{t}c$.
For instance, for Higgs mass heavier than 250 GeV, we get a branching ratio
of the order $10^{-14} \to 10^{-12}$  for $H\to {\bar{t}}c$.
New sources of flavor violation like charged Higgs are expected to
 enhance the branching ratio of $\Phi\to \bar{t}c$.

\subsection{$\Phi \to {\bar{t}}c$}
In this section we discuss the 2HDM contributions to the 
Higgs FCNC.
We concentrate mainly on $h^0\to \bar{t}c$, $H^0\to \bar{t}c$ and 
$A^0\to \bar{t}c$. 
The generic Feynman diagrams which contribute to
those FCNC are shown in fig~.(\ref{hsbb}).
As we can see, the additional interactions like $\Phi^0H^+H^-$,
$\Phi^0H^+G^-$ and $\Phi^0H^+W^-$ give new contributions to
$\Phi^0\to \bar{t}c$.

The calculation of $h^0\to \bar{t}c$, $H^0\to \bar{t}c$
and $A^0\to \bar{t}c$ in 2HDM has been first done in Ref.~\cite{2hdm4}.
In Ref.~\cite{2hdm4}, only trilinear Higgs self-couplings
$\lambda_{HHH}$ have been 
constrained by the maximum unitarity limit tolerated for the SM
trilinear coupling: 
\begin{eqnarray}
|\lambda_{HHH}| \leq |\lambda_{HHH}^{SM}(M_H=1 {\rm TeV})|=\frac{3 g (1
 {\rm TeV})^2}{2 M_W}\label{unitt}
\end{eqnarray}
We have checked that the above constraints (\ref{unitt}) 
are not enough to guarantee unitarity and perturbativity constraints.
For instance, for the parameter set considered in Ref.~\cite{2hdm4},
$\lambda_1$ can take extremely large values of the order $\approx 5000$
which is not tolerated neither by perturbativity nor by 
unitarity constraints. We perform a cross check with 
Ref.~\cite{2hdm4}, and our results\footnote{We reproduce our results by 2
different and independent programs.} are about 3 times 
larger than those presented in Ref.~\cite{2hdm4}. 

The decay width $\Gamma_{\Phi}$ of all scalar particles 
$h^0$, $H^0$ and $A^0$ are computed at tree level in 2HDM as follows: 
\begin{eqnarray}
\Gamma_{\Phi}=\sum_{f}\Gamma(\Phi\to f\bar{f}) + 
\Gamma(\Phi\to VV)  +  \Gamma(\Phi\to V H_i)
+\Gamma(\Phi\to H_i H_j)\label{widd}
\end{eqnarray}
QCD corrections 
to $\Phi \to f\bar{f}$ and  $\Phi \to \{ g g, \gamma \gamma, \gamma Z,
V^*V^*, VV^*, V^*H_i\}$ decays are not included in the width.
Since the width of $\Phi\to \bar{t}c$ does not become
comparable to $\Phi \to \{ g g, \gamma \gamma, \gamma Z\}$,
it is not necessary to include them in the computation.
The full width of the Higgs bosons is taken from
\cite{AKZ} without QCD corrections. 
\begin{figure}[t!]
\smallskip\smallskip 
\vskip-.1cm
\centerline{{
\hskip-.04cm
\epsfxsize3.18 in 
\epsffile{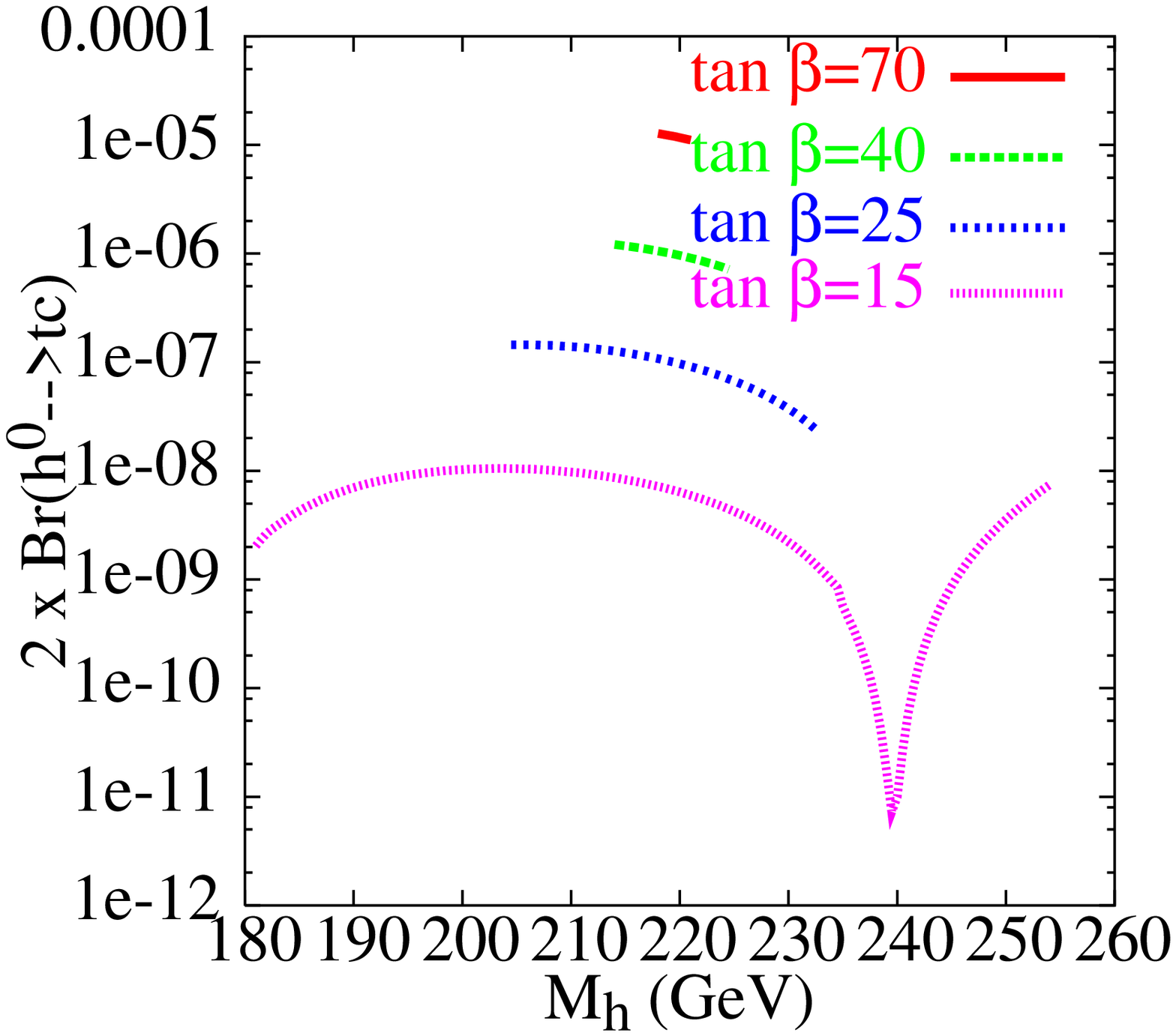}} 
\hskip-.01cm
\epsfxsize3.18 in 
\epsffile{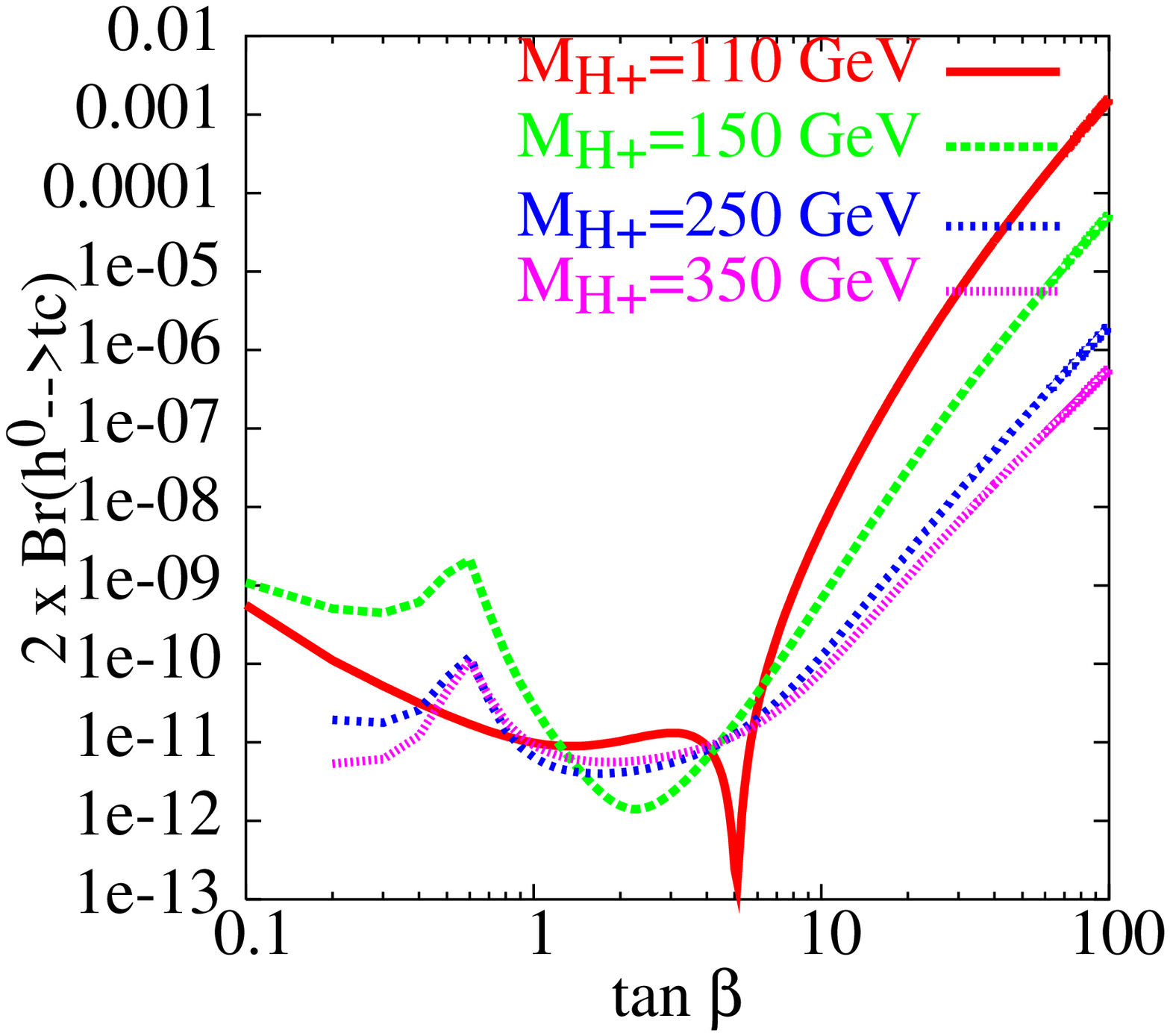}}
\smallskip\smallskip
\caption{$2\times Br(h^0\to \bar{t}c)$ in 2HDM-II
as a function of $M_h$ (left) and $\tan\beta$ (right) with
$M_{A^0}=215$, $M_{H^0}=250$ GeV, $\lambda_5=2$ and 
$\sin\alpha=0.5$ and $M_{H\pm}=150$ GeV (left) and 
$M_{h^0}=220$ GeV(right)}
\label{fig7}
\end{figure}
\begin{figure}[t!]
\smallskip\smallskip 
\vskip.1cm
\centerline{{
\epsfxsize3.18 in 
\epsffile{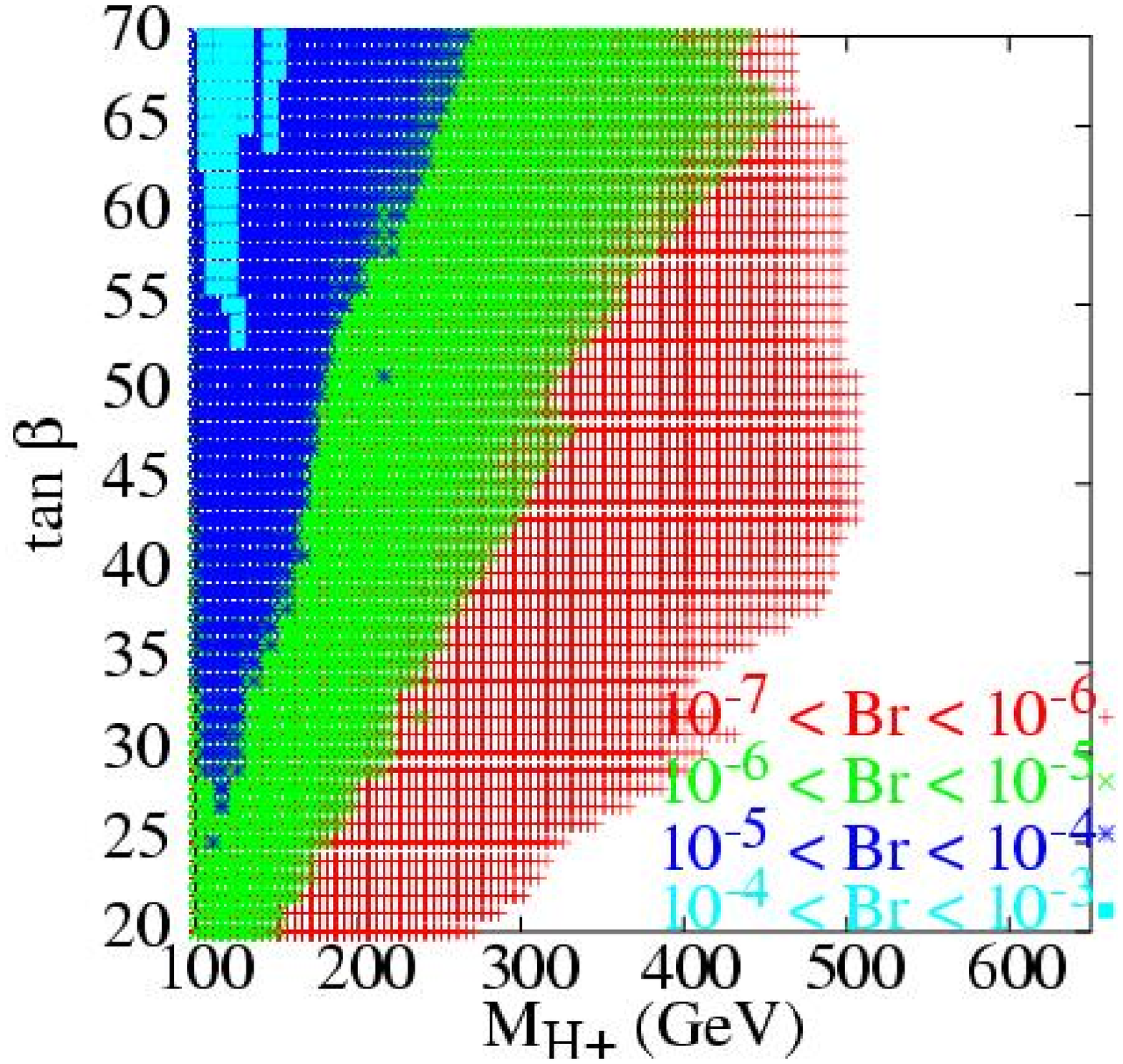}}  
\hskip-.01cm
\epsfxsize3.1 in 
\epsffile{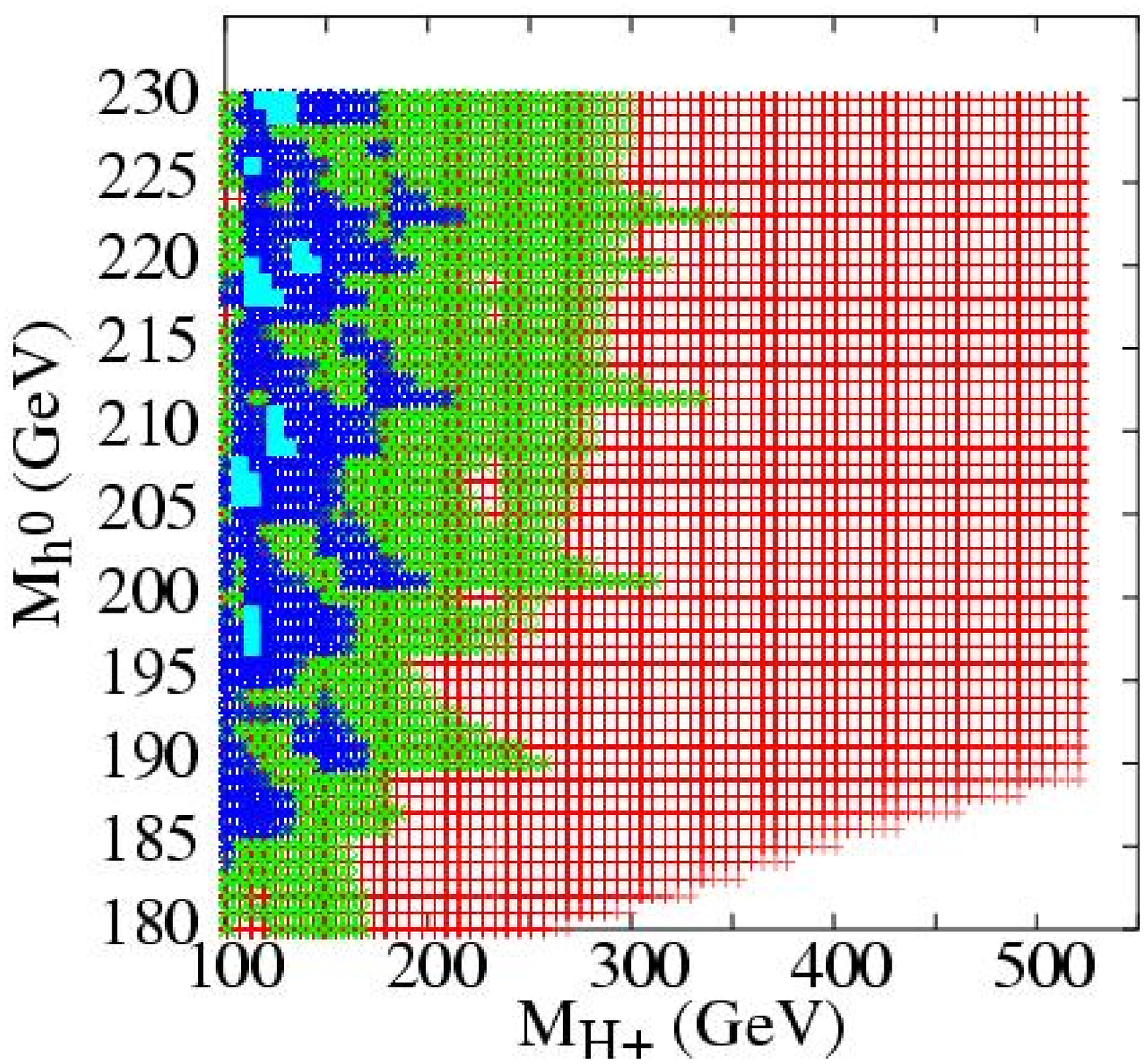 }}
\smallskip\smallskip
\caption{Scatter plots for $2 \times {\rm Br}(h^0\to \bar{t}c)$ in the 
  ($\tan\beta$, $M_{H\pm}$) (left) and ($M_{h^0}$, $M_{H\pm}$)
  plans. We scan over: $180 \la M_{h^0} \la 230$, $
190 \la M_{H^0} \la 230$
$100 \la M_{A^0,H\pm} \la 600$ GeV, $-1\la \sin \alpha \la 1$ and 
$20\la \tan\beta \la 70$ and $\lambda_5=M_A^2/v^2$.}
\label{fig77}
\end{figure}
As stated before, we perform a systematic scan over the 
2HDM parameter space using  unitarity constraints and vacuum stability
conditions on the scalar sector parameters. 
In our numerical analysis, we limit ourself to 2HDM type II.
As shown in Ref.~\cite{2hdm4}, the rates in 2HDM-I are smaller 
than the rates in 2HDM-II. We also present numerics for CP even Higgs
only  $\Phi=h^0, H^0$ and comment on our finding for 
$A^0\to \bar{t}c$ in 2HDM type I and II.

\subsubsection*{$h^0\to \bar{t}c$}
We illustrate first in Fig.~(\ref{fig7}) the branching ratio of 
$h^0\to \bar{t}c$ in 2HDM-II as a function of $M_{h^0}$ (left) and $\tan\beta$
(right). The parameters are fixed as:
$M_{A^0}=215$, $M_{H^0}=250$ GeV, $\lambda_5=2$ and 
$\sin\alpha=0.5$ and $M_{H\pm}=150$ GeV (left) and $M_{h^0}=220$ 
GeV (right). In both plots (left and right), one can see that 
the Branching ratio of $h^0\to \bar{t}c$, like the Br$(t\to ch^0)$, 
can be of the order $10^{-5}$ 
only for large $\tan\beta$ and relatively light charged Higgs mass.
As in the case of $t\to c h^0$, this enhancement at large $\tan\beta$ 
 is due to bottom Yukawa coupling as well as to the triple scalar
coupling $h^0H^+H^-$.\\
We show on the right panel of Fig.~(\ref{fig7}) 
the Branching ratio of $h^0\to \bar{t}c$ as a function of $\tan\beta$ for 
several values of charged Higgs mass $M_{H\pm}$.
The maximum rate for Br($h^0\to \bar{t}c)\approx 4\times 10^{-4}$ 
is obtained for 
$M_{H\pm}=110$ GeV and
large  $\tan\beta\approx 65$. 
The reason for this enhancement is due 
to threshold effect, since at $M_{h^0}=220$ GeV and  $M_{H\pm}=110$ GeV
the decay channel $h^0\to H^+H^-$ is open\footnote{see next plot for
  similar threshold effect for $H^0\to \bar{t}c$}. 
For charged Higgs mass away from the
threshold region, the Br($h^0\to \bar{t}c$) is reduced and for 
$M_{H\pm}=150$ GeV, it is of the order $10^{-5}$ at $\tan\beta\approx 70$.
\\
In the scatter plots Fig.~(\ref{fig77}) it is illustrated
in the left panel that for large $\tan\beta\ga 30$
and light charged Higgs $M_{H\pm}\la 250$ GeV the 
${\rm Br}(h^0\to \bar{t}c)\in[10^{-5},10^{-3}]$. The maximum 
branching ratio is obtained close to threshold region where 
the decay $h^0\to H^+H^-$ is open.
For charged Higgs mass $M_{H\pm}\ga 250 $ GeV, ${\rm Br}(h^0\to
\bar{t}c)\la 10^{-5}$ for all range of $\tan\beta$.  
From the right panel, one can see that the ${\rm Br}(h^0\to \bar{t}c)$ 
is not very sensitive to the CP-even mass $M_{h^0}$.

\begin{figure}[t!]
\smallskip\smallskip 
\vskip.1cm
\centerline{{
\epsfxsize3.18 in 
\epsffile{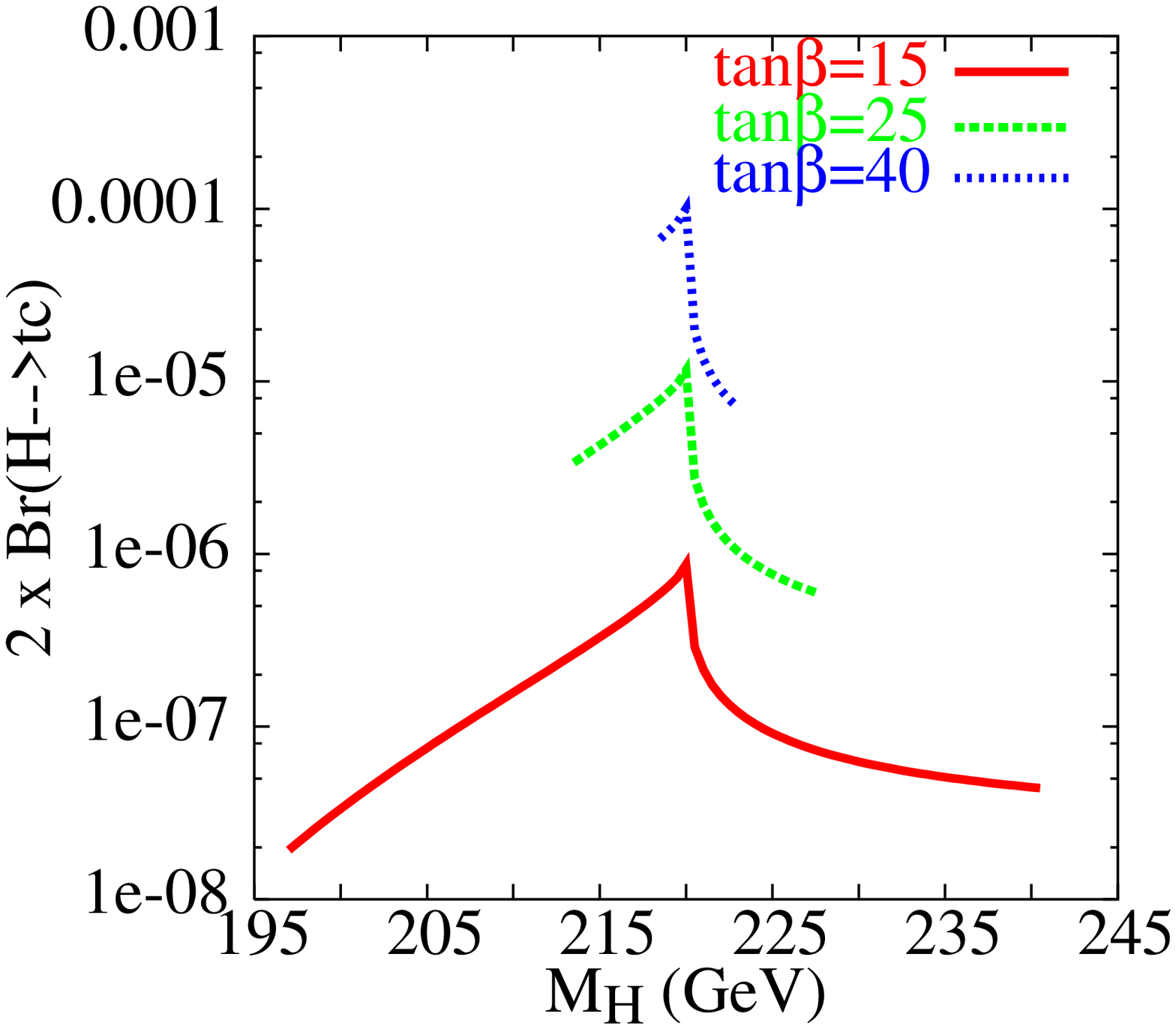}}  
\hskip-.01cm
\epsfxsize3.1 in 
\epsffile{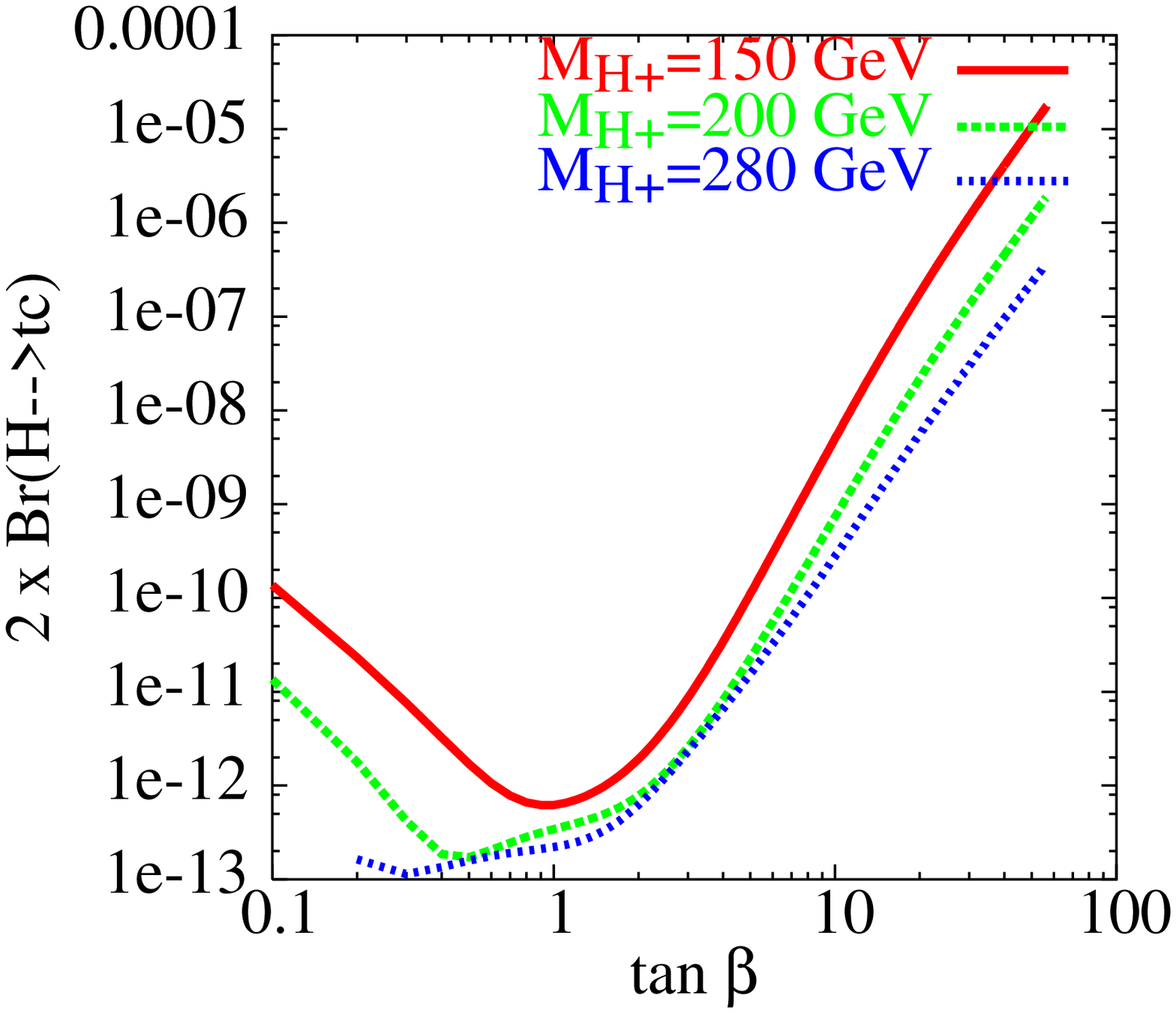}}
\smallskip\smallskip
\caption{Branching ratio for $2\times Br(H^0\to \bar{t}c)$ as
 a function of $M_{H^0}$ (left) and  $\tan\beta$ (right) in 2HDM-II 
$M_{h^0}=120$, and $M_{A^0}=200$ GeV,   
$\lambda_5=1$, $\sin\alpha=0.75$ with $M_{H\pm}=110$ GeV for left and
$M_{H^0}=220$ GeV for right panel}
\label{fig8}
\end{figure}
\begin{figure}[t!]
\smallskip\smallskip 
\vskip.1cm
\centerline{{
\epsfxsize3.18 in 
\epsffile{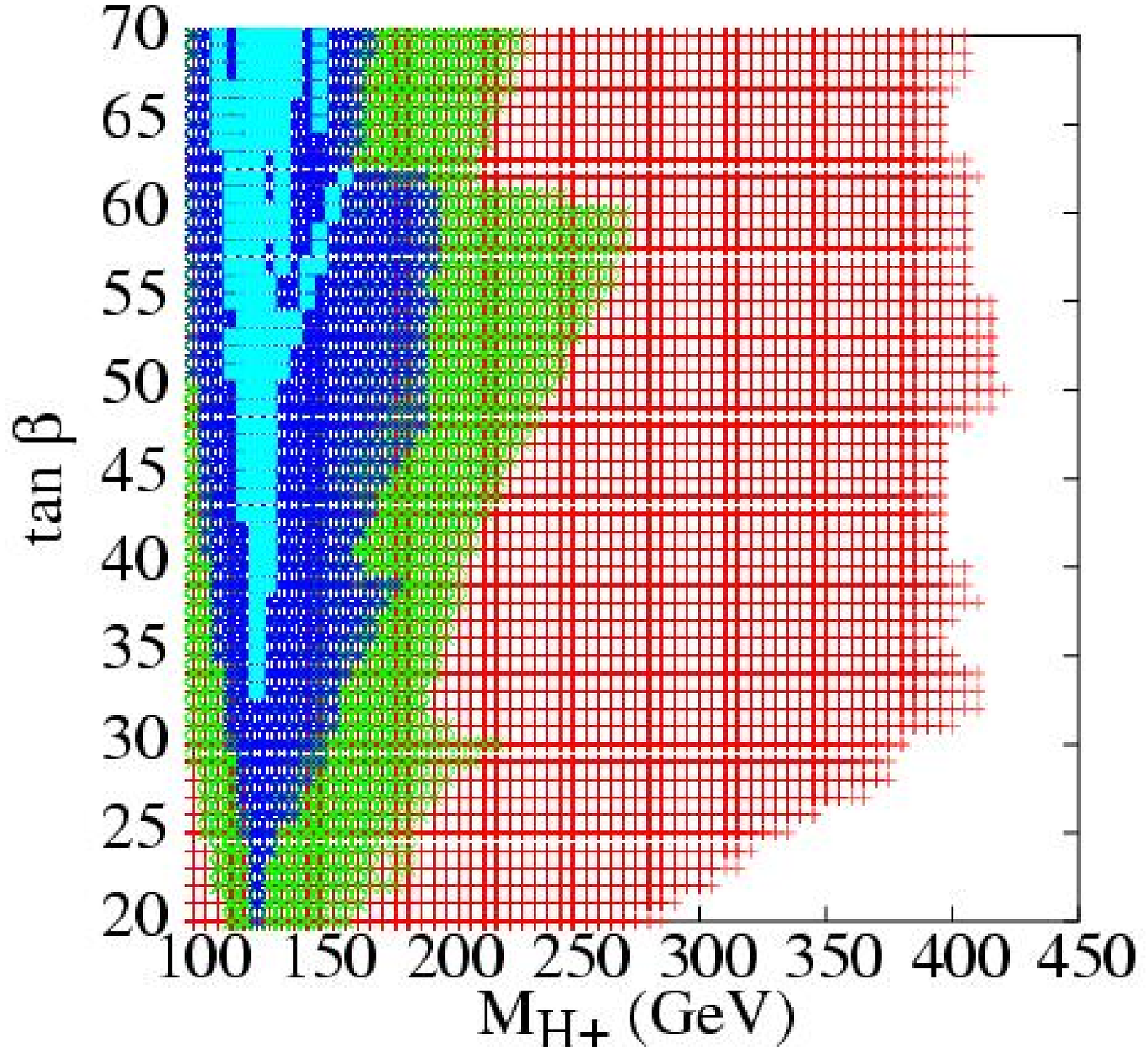}}  
\hskip-.01cm
\epsfxsize3.1 in 
\epsffile{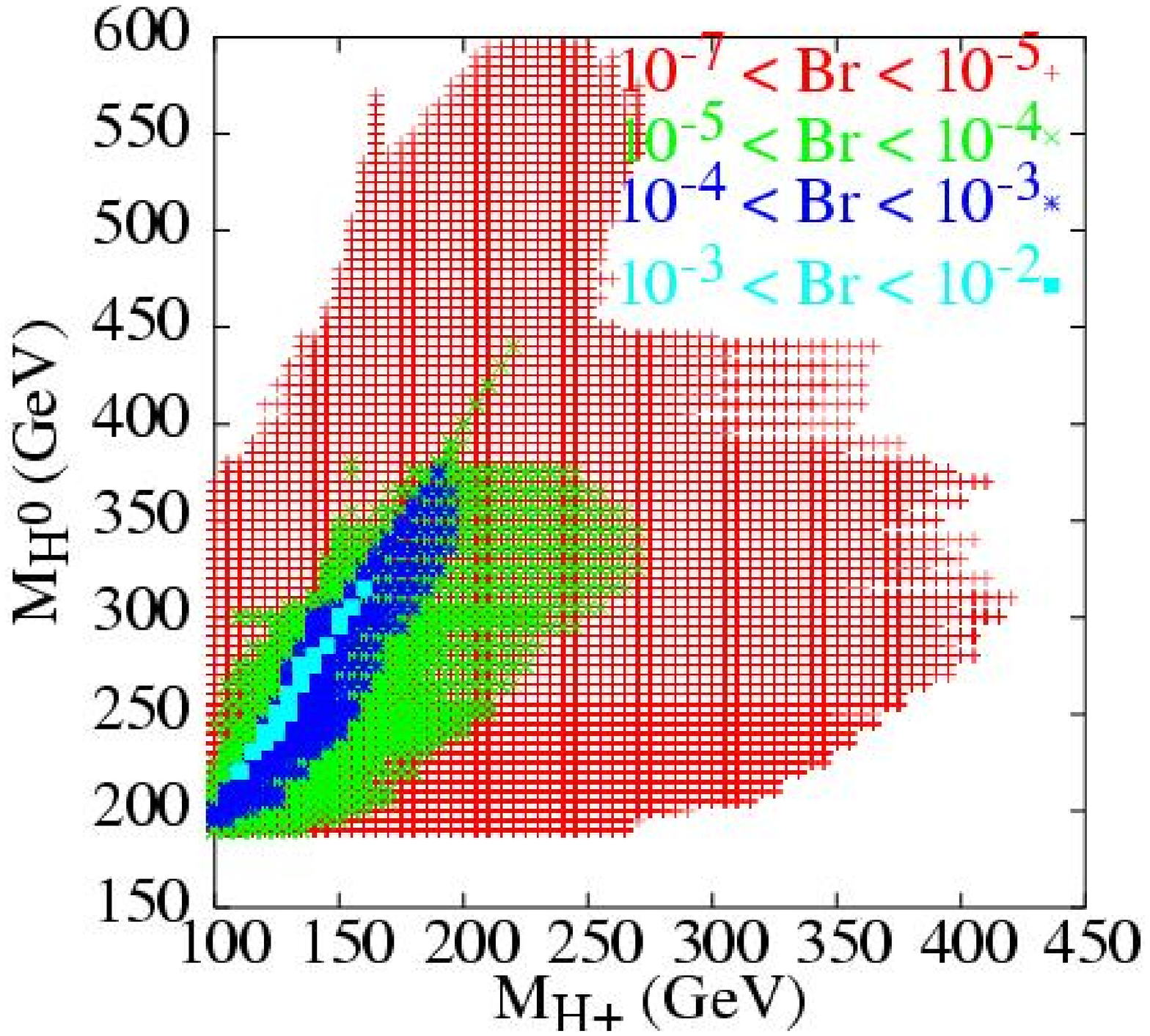 }}
\smallskip\smallskip
\caption{Scatter plots for $2 \times {\rm Br}(H^0\to \bar{t}c)$ 
in the  ($\tan\beta$, $M_{H\pm}$) (left) and ($M_{H^0}$, $M_{H\pm}$)
  plans. We scan over: $100 \la M_{h^0} \la 230$, $
180 \la M_{H^0} \la 600$
$100 \la M_{A^0,H\pm} \la 600$ GeV, $-1\la \sin \alpha \la 1$ and 
$20\la \tan\beta \la 70$ and $\lambda_5=M_A^2/v^2$.}
\label{fig88}
\end{figure}

\subsubsection*{$H^0\to \bar{t}c$}
We now discuss the 2HDM contribution to $H^0\to \bar{t}c$.
We perform a scan in the 2HDM parameter space and find 
a region where the branching ratio is sizeable. The conclusion is
similar to $h^0\to \bar{t}c$. To have a large branching ratio we need
to maximize the coupling $H^0H^+H^-$ and the bottom Yukawa coupling.
This is achieved by taking large $\tan\beta$.\\
In the left panel of Fig.~(\ref{fig8}), we illustrate
the Branching ratio of $H^0\to \bar{t}c$ as a function of CP 
even Higgs mass $M_H$ for several values of $\tan\beta$. 
As it can be seen, for moderate $\tan\beta <15$,
the Branching ratio is less than $10^{-6}$. When the CP even mass
$M_{H^0} \approx 2 M_{H\pm} \approx 220$ GeV one can see a  
peak. Such a peak corresponds to the opening of the decay 
channel $H^0\to H^+H^-$.
This threshold effect is also amplified by the fact that the 
coupling $H^0H^+H^-$ takes its maximal value for 
$M_{H^0} \approx 220$ GeV and large $\tan\beta$.  
In this region, the Branching ratio can reach 
$3\times 10^{-4}$  at large $\tan\beta\approx 50$.
As can be seen in the left plot, 
for large $\tan\beta\geq 25$ unitarity is violated apart from 
a small region around $M_H\approx 220$ GeV. 
Once the Higgs mass $M_H$ becomes larger than $2M_{H\pm}$,
the decays $H^0\to \{H^+H^-,h^0h^0\}$ are open, 
the width $\Gamma_{H^0}$
becomes large and hence the Branching ratio is reduced.

In the right panel of Fig.~(\ref{fig8}), we show the
branching ratio for $H^0\to \bar{t}c$ as a
function of $\tan\beta$ for several values of $M_{H\pm}$.
As for the $h^0\to \bar{t}c$ case, the branching ratio is enhanced by
large $\tan\beta$ and rather light charged Higgs mass.
Also in this case, for large $\tan\beta$ the 
coupling $H^0H^+H^-$ reaches its maximal value.
It is clear from the right plot that the Branching ratio
$Br(H^0\to \bar{t}c)$ can be greater than $10^{-5}$ only 
for large $\tan\beta\ga 45$ and $M_{H\pm}=150$ GeV.
Once the charged Higgs mass becomes large, the $Br(H^0\to
\bar{t}c)$ decreases. This can be seen in the right
plot of Fig.~(\ref{fig8}) where $M_{H\pm}=250$ GeV,
the Br$(H^0\to \bar{t}c)$ is less than $10^{-6}$.\\
We have also performed a systematic scan for this case, and 
the results are illustrated in Fig.~\ref{fig88}. 
In those scatter plots, we show ${\rm Br}(H^0\to \bar{t}c)$
as function of  ($M_{H\pm}$,$\tan\beta$) (left) 
and ($M_{H\pm}$,$M_{H^0}$) (right).
Again, close to the threshold  region where $H^0\to H^+H^-$ is open,
the ${\rm Br}(H^0\to \bar{t}c)$ can be larger than $10^{-3}$ (see left
panel of Fig.~\ref{fig88}). 
Such an enhancement requires of course $M_{H\pm}\la 150$ GeV 
and large $\tan\beta$. In this case also, there is a region in which 
the charged Higgs mass is larger than 200 GeV, which could accommodate
$B\to X_s\gamma$, and the ${\rm Br}(H^0\to \bar{t}c)\in [10^{-5},10^{-4}]$.
From the right panel of Fig.~\ref{fig88}, one can see that once the 
$\bar{t}t$ threshold has been passed $M_H\ga 350$ GeV, 
the Branching ratio ${\rm Br}(H^0\to \bar{t}c)$ is reduced.

\subsubsection*{$A^0 \to {\bar{t}}c$}
At the end of this section, we would like to comment 
on $A^0\to \bar{t}c$ decay in 2HDM 
type I and II. In 2HDM-II, we perform a systematic scan 
over the 2HDM parameters. It is found that the Branching ratio of 
$A^0\to \bar{t}c$ decay never exceeds $10^{-8}$ (resp $3 \times 10^{-9}$) for
small (resp large) $\tan\beta$. The reason is that, unlike $h^0$ and
$H^0$, the CP-odd $A^0$, being pseudo-scalar, 
does not couple to a pair of charged Higgs. Therefore $A^0\to
\bar{t}c$, has only a unique source of enhancement which is the bottom
Yukawa coupling.

In 2HDM-I, there is a small region
for $100 < M_{H\pm} < 200 $ GeV, $176 < M_{A^0} < 2 M_t\approx 350 $ GeV
and small $0.1 <\tan\beta < 0.3$ where the Branching ratio of 
$A^0\to \bar{t}c$ is in the range: $[10^{-9},3\times 10^{-7}]$. 
At large $\tan\beta$ the branching ratio decreases
quickly and is less than $10^{-14}$ for $\tan\beta > 2$.
The origin of this is the fact that the couplings
$A^0b\bar{b}$ and $H^-t\bar{b}$ are proportional to $1/\tan\beta$
and so suppressed for large $\tan\beta$.

\section{Top-charm production at 
$e^+e^-$, $\gamma\gamma$ and muon colliders}
\begin{figure}[t!]
\begin{center}
\vspace{-2.cm}
\input{ee.tex}
\vspace{-12.99cm}
\caption{Topological contributions to $e^+e^-\to \bar{t}c$ and
  $\mu^+\mu^-\to \bar{t}c$ in 2HDM }
\label{eee}
\end{center}
\end{figure}
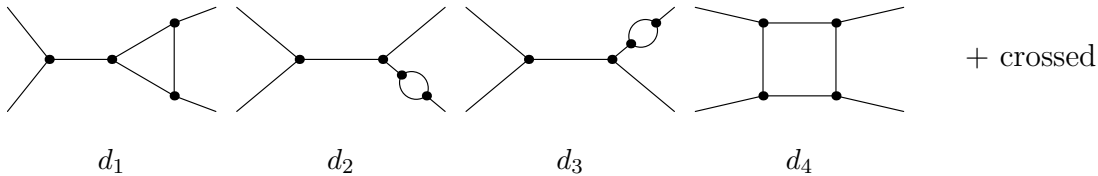
\begin{figure}[t!]
\begin{center}
\vspace{-3.cm}
\input{gg.tex}
\vspace{-9.99cm}
\caption{Topological contributions to $\gamma\gamma\to \bar{t}c$ in 2HDM}
\label{ggg}
\end{center}
\end{figure}
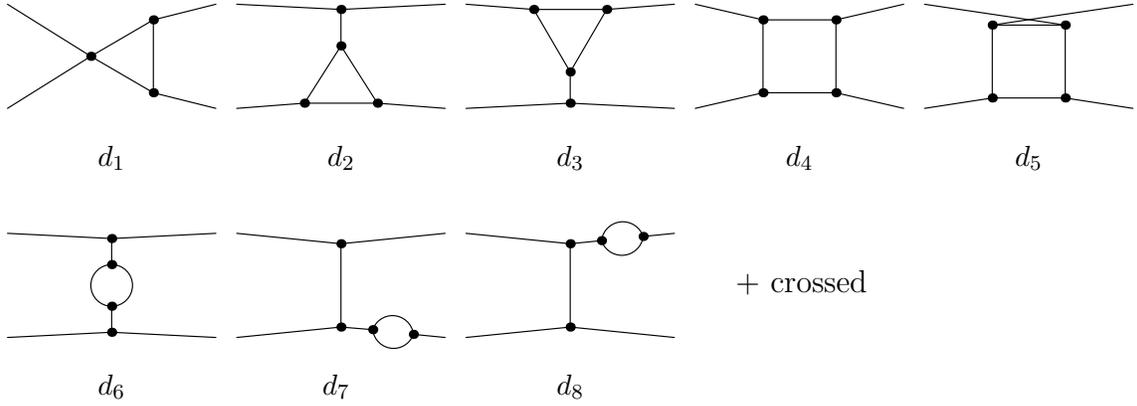
There have been several studies looking for collider signatures
of the top and Higgs FCNC couplings both at lepton colliders
\cite{soni} as well as at hadron colliders \cite{sola}. 
Because of the very large mass of the top, 
Top-charm production at leptons and hadron colliders would have 
a clear signature.\\ 
At LHC, the production and subsequent decay of the neutral CP-even Higgs 
mediated by FCNC have been analyzed in \cite{2hdm4} for 2HDM and
\cite{sola} for MSSM. It has been demonstrated in those studies
that the production rates for the $\bar{b}s+b\bar{s}$ channel
can be large while the rates for $\bar{t}c+t\bar{c}$ channel 
are more modest. Detection for the former channel at LHC is 
difficult but it is easier for the later \cite{sola}.\\
Associate Top-charm production at $e^+e^-$ colliders 
has been studied in SM \cite{9901369} and found to be very small.
New physics contribution to $e^+e^-\to \bar{t}c$ 
such as SUSY can enhance the cross sections
by several orders of magnitude with respect to SM \cite{9904273}.
The top-charm production at lepton colliders $e^+e^-$ and 
$\mu^+\mu^-$ has been studied in two Higgs 
doublet models without Natural Flavor Conservation 2HDM-III 
which can lead to measurable effects \cite{soni,sonimu}.  
It has been pointed out that the tree level vertex $\Phi \bar{t}c$ can be
better probed through t-channel W-W and/or Z-Z fusion at 
high energy $e^+e^-$ collisions
$e^+e^- \to t \bar{c}\nu_2 \bar{\nu}_e$ and 
$e^+e^- \to t \bar{c} e^+ e^-$ \cite{2hdm11,hou}.
An interesting feature of these reactions is that, being t-channel fusion
processes, their cross sections grow with energy unlike
s-channel reactions $e^+e^- \to t \bar{c}$ 
which are suppressed at high energies. 
The cross sections of 
$e^+e^- \to t \bar{c}\nu_e \bar{\nu}_e$ and 
$e^+e^- \to t \bar{c} e^+ e^-$ are found to be 
one or two orders of magnitude bigger than the cross sections of 
$e^+e^- \to t \bar{c}$ \cite{2hdm11,hou}.\\
In the present study, we limit ourself to $e^+e^-$ colliders, its
$\gamma\gamma$ option and muon colliders.
The  cross sections for $e^+e^- \to  \bar{t}c$,  
$\gamma\gamma \to \bar{t}c$ and/or  
$\mu^+\mu^- \to  \bar{t}c$ in 2HDM, if sizeable,
 can give information on the top FCNC couplings 
$t\to c\gamma $ and $t \to c Z$ as well as Higgs FCNC $\Phi\to \bar{t}c$.
Unlike the process $e^+e^- \to \bar{t}c$, which is s-channel suppressed
at high energy, the process $\gamma\gamma \to \bar{t}c$ has a t and
u-channel contributions, its cross section may grow with 
the center of mass energy.\\
At $e^+e^-$ collisions, the generic contributions to 
$e^+e^-\to \bar{t}c$ are drawn in Fig.~(\ref{eee}). We have three 
sets of diagrams: $\gamma \bar{t}c $ vertex contributions 
through $e^+e^- \to \gamma^*\to
\bar{t}c$, $Z \bar{t}c$ vertex contributions through $e^+e^- \to Z^*\to
\bar{t}c$  and box contributions. The topological contributions
to $e^+e^- \to   \bar{t}c$ are depicted in Fig~.(\ref{eee}).\\
In $\gamma\gamma$ collisions, we have more topological diagrams
that contribute to $\gamma \gamma \to \bar{t}c$. $\gamma \bar{t}c$ vertex
enters game through diagrams $d_2$, $d_3$, $d_7$ and $d_8$
 of Fig.~(\ref{ggg}) and their crossed one. The quartic coupling of 2
 photons to a pair of $H^+H^-$ or $W^+W^-$ gives the
 topological diagram $d_1$. Box (resp self-energies) topologies 
are depicted in diagrams $d_4$ and $d_5$ (resp $d_6$). We have
checked both analytically and numerically that the result is 
Ultra-Violet finite and renormalization scale independent.
In the case of $\gamma\gamma \to  \bar{t}c$, gauge invariance 
has been checked numerically.

\begin{figure}[t!]
\smallskip\smallskip 
\vskip-.01cm
\centerline{{
\epsfxsize3.1 in 
\epsffile{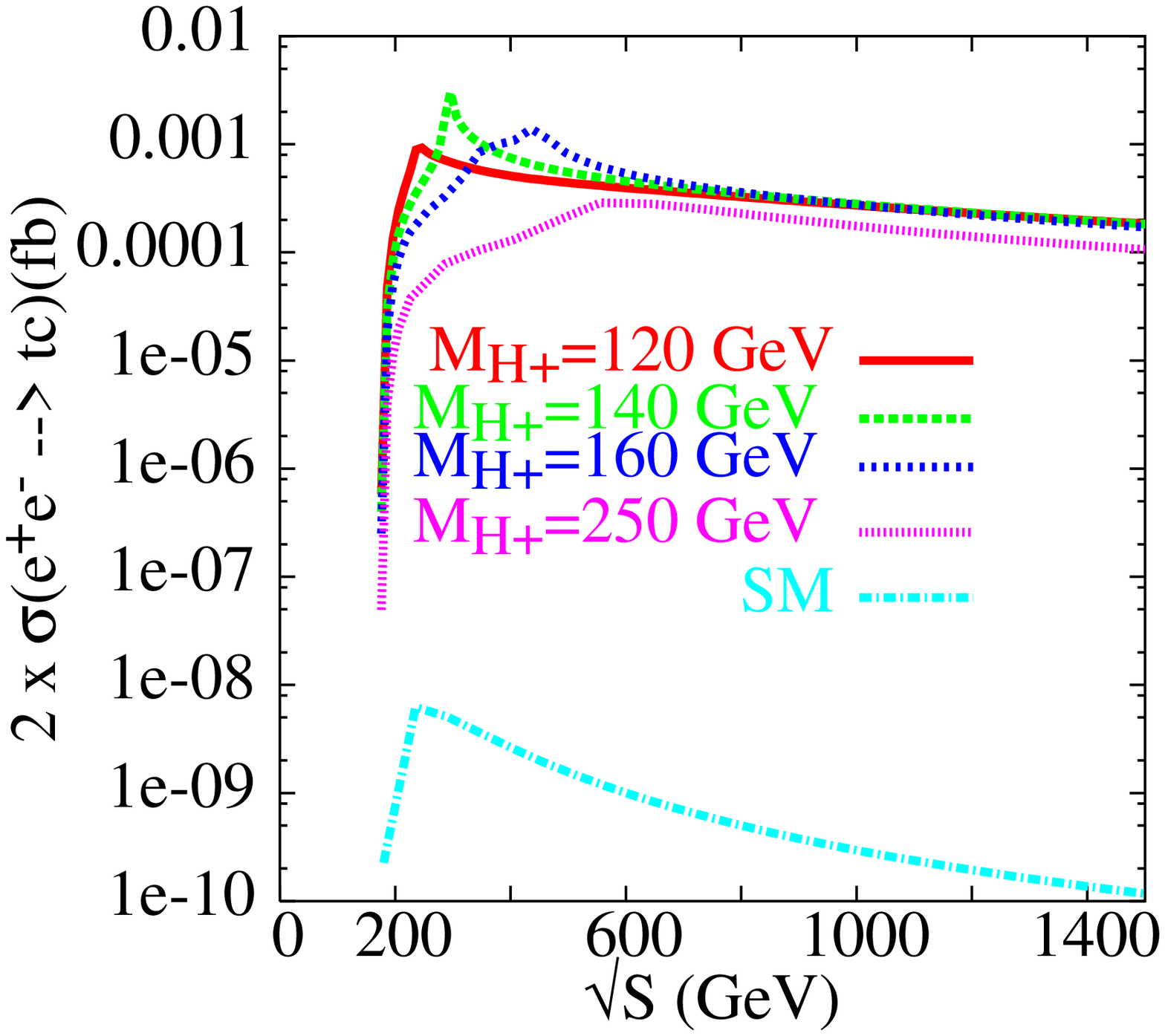}}  \hskip0.4cm
\epsfxsize3.1 in 
\epsffile{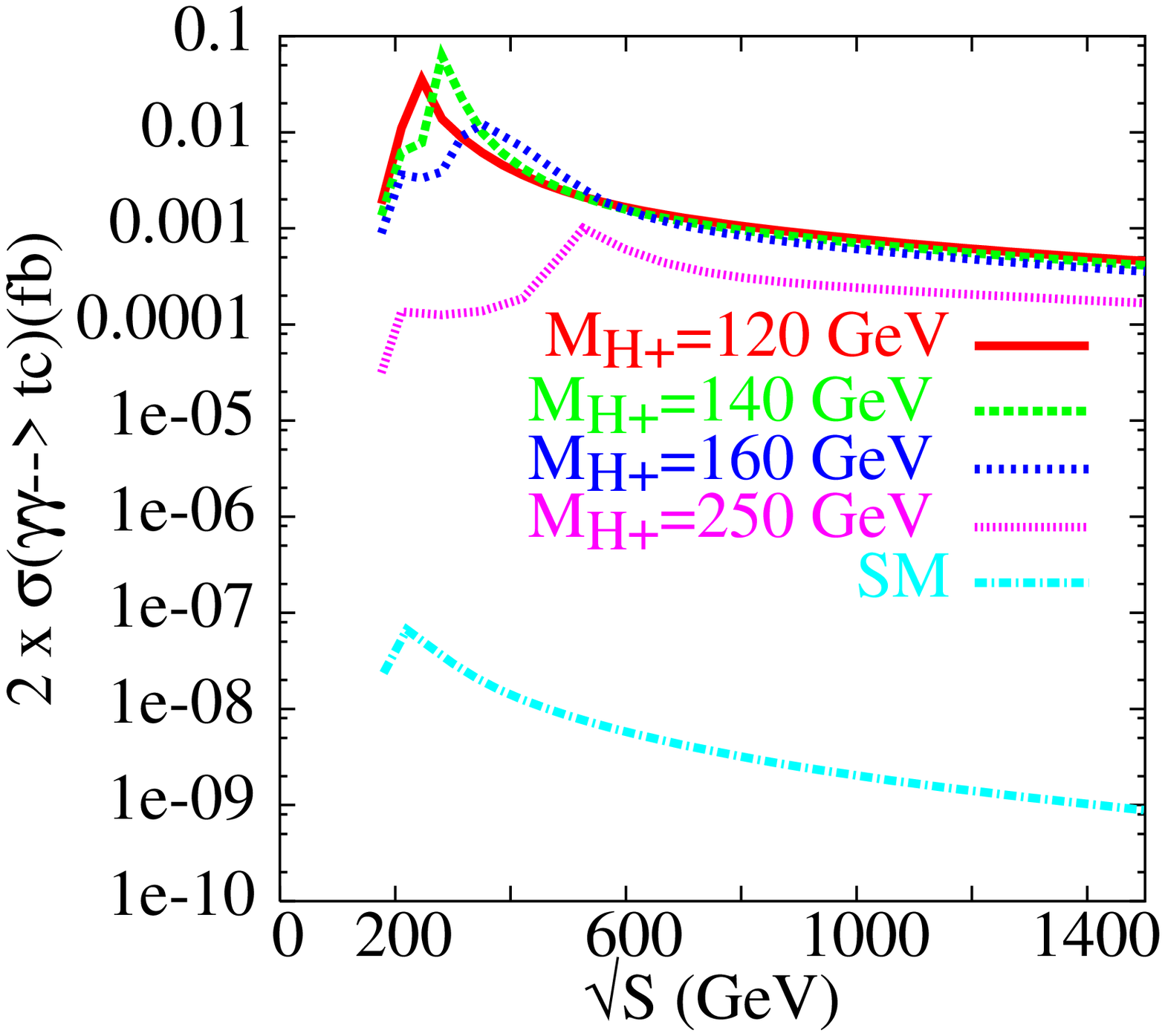} }
\smallskip\smallskip
\caption{Cross sections for associated top-charm production
at $e^+e^-$ colliders (left) and its $\gamma \gamma$ option (right)
in 2HDM-II with $\tan\beta=65$ and several choices of charged Higgs mass}
\label{fig13}
\end{figure}

Due to GIM mechanism, in the SM, the cross sections of 
$e^+e^- \to \bar{t}c$ and $\gamma\gamma \to \bar{t}c$
are very suppressed see Fig.~(\ref{fig13}) (left and right).
It is clear from Fig.~(\ref{fig13}) that, in the SM, the 
$\gamma\gamma \to \bar{t}c$  cross section is about one order 
of magnitude bigger than $e^+e^- \to \bar{t}c$ cross sections. 
Our results for the SM agree with \cite{9901369}\\
We illustrate our numerical results in 2HDM-II for large 
$\tan\beta = 65$ and several values of charged Higgs.
The reason for this choice is the fact that the one loop effective
coupling $t\to c \gamma$ and $t \to c Z$ are larger 
for large $\tan\beta$ (see Fig.~(\ref{fig5})).
We consider only unpolarized cross sections. It is well known that
the cross sections for polarized initial states differ from the
unpolarized cross sections only by the normalization factor.

In both cases, for $e^+e^-$ and $\gamma\gamma$ collisions,
the cross sections in the 2HDM are enhanced by 
about four orders of magnitude with respect to their SM values.
Above $\sqrt{s}=200$ GeV, the cross sections become sizeable and 
develop a peak at $\sqrt{s}=2 M_{H\pm}$, which is due to the
threshold effect of  charged Higgs pair production. 
The cross sections can reach
a value of $0.003$ fb (resp 0.1 fb) for $e^+e^- \to \bar{t}c$
(resp $\gamma\gamma \to \bar{t}c$) near threshold.
At $\gamma \gamma $ options, cross sections of about 
$0.01\to 0.1$ fb would give $4\to 80$ events 
for the assumed luminosities of $400\to 800$ fb$^{-1}$.
Above the threshold region $\sqrt{s}\ga 800$ GeV,
the cross section decreases with increasing the energy
and reaches a value of $\approx 10^{-4}$ fb at $\sqrt{s}\ga 1.5$ TeV.\\
In the 2HDM-I, the cross sections
are below $\approx 10^{-3}$ fb (resp $10^{-5}$ fb) for $\gamma\gamma$
(resp $e^+e^-$) collisions.
\begin{figure}[t!]
\smallskip\smallskip 
\vskip-.01cm
\centerline{{
\epsfxsize3.1 in 
\epsffile{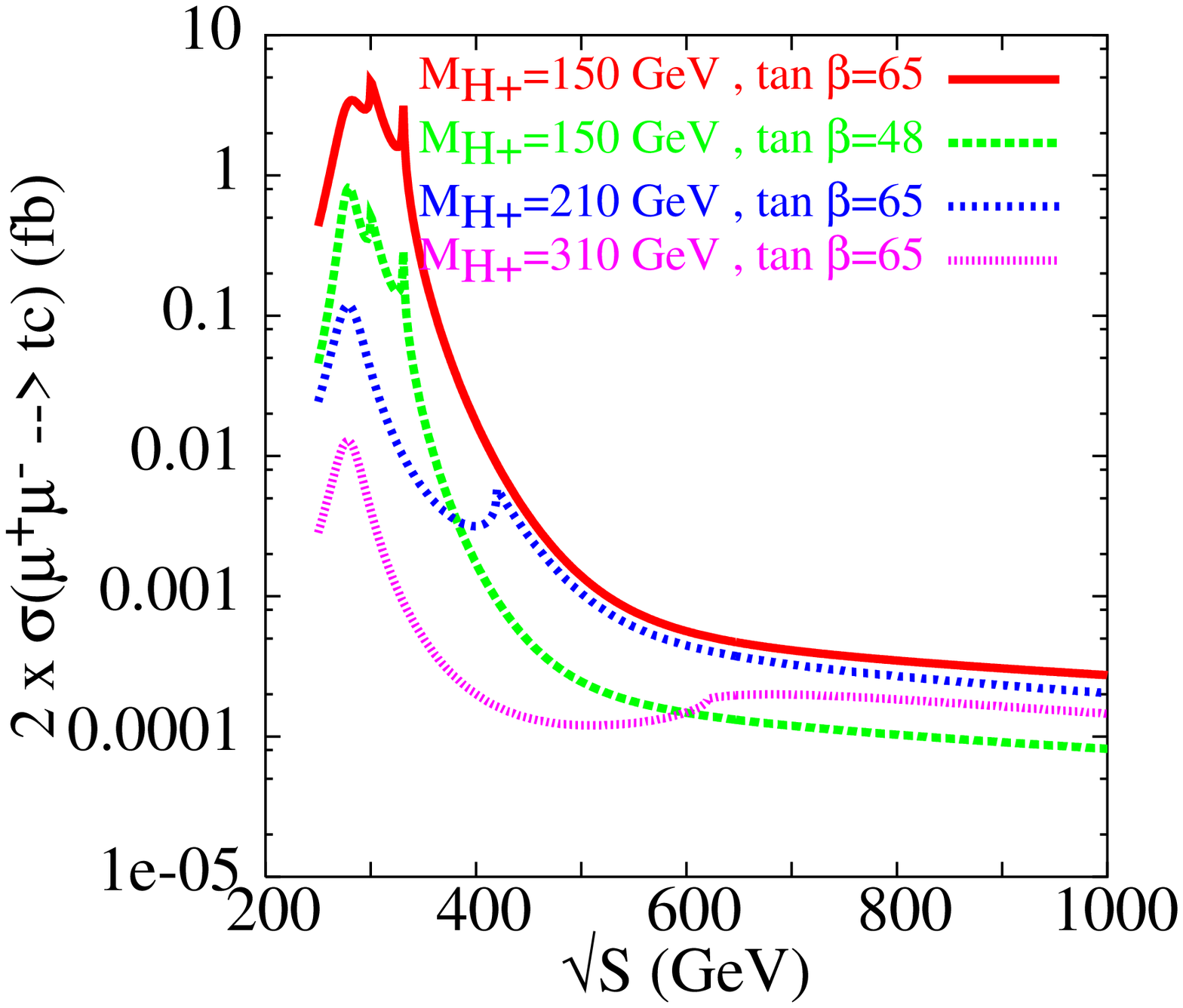 }}  \hskip0.4cm
\hskip0.4cm
\epsfxsize3.1 in 
\epsffile{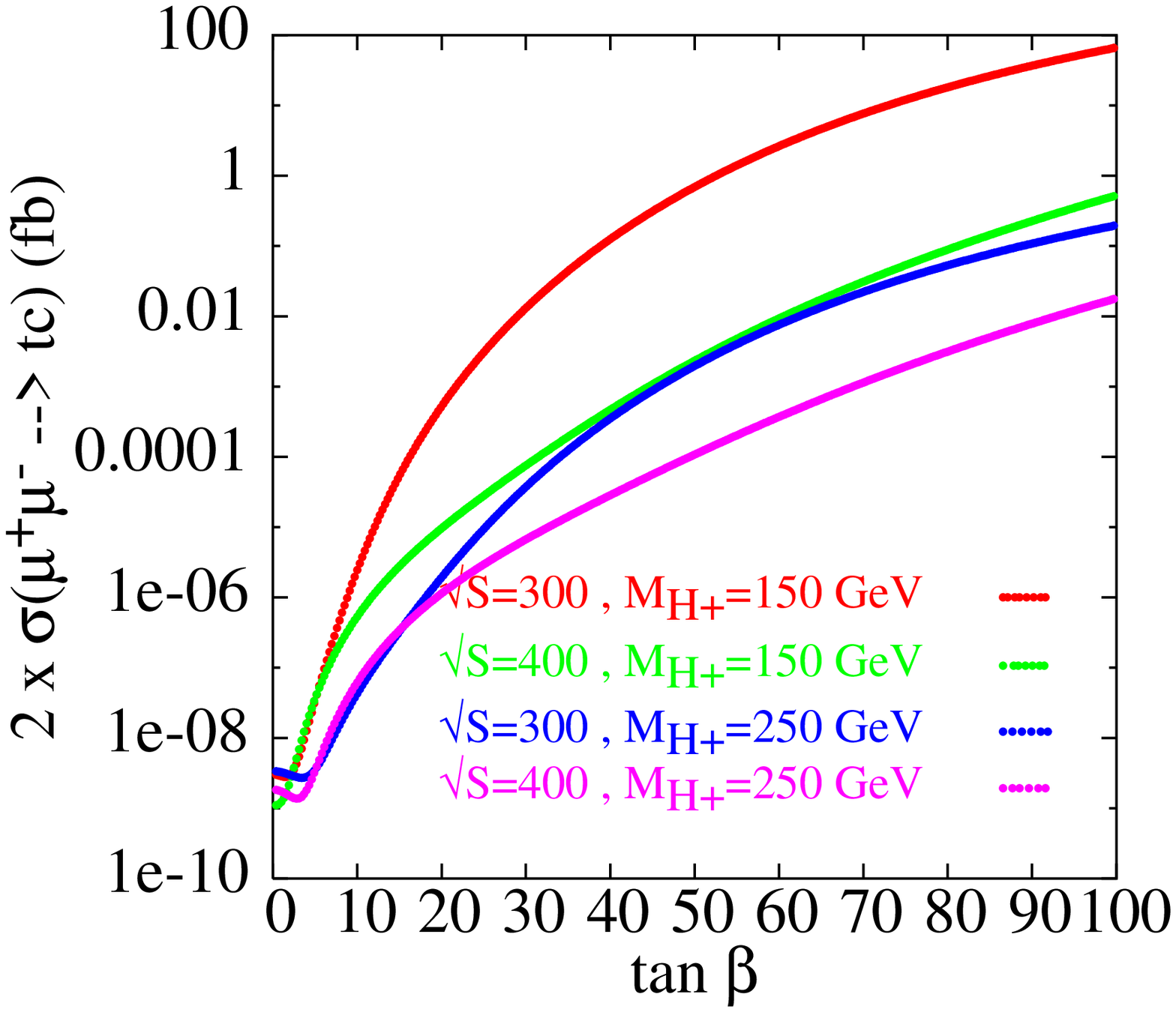} }
\smallskip\smallskip
\caption{ $\mu^+\mu^-\to t\bar{c}+\bar{t}c$ cross section
as function of $\sqrt{s}$ (left) and $\tan\beta$ (right)
for $M_h=100$,$M_H=278$, $M_A=170$ GeV, $\sin\alpha=-0.84$
and $\lambda_5=M_A^2/v^2=0.95$.}
\label{muu}
\end{figure}

In recent years an increasing amount of work has been dedicated to
the physics possibilities of $\mu^+\mu^-$ colliders
\cite{Muon}. Such colliders
offer novel ways of producing Higgs bosons, such as an $s$--channel
resonance in the case of neutral scalars.
The production process $\mu^+\mu^- \to \bar{t}c$ proceed through
the topological diagrams drawn in Fig.~\ref{eee}.
As for $e^+e^-\to \bar{t}c$ we have the gauge boson 
s-channel $\mu^+\mu^- \to
\gamma,Z^*\to \bar{t}c$
contributions as well as the Higgs boson s-channel exchange 
$\mu^+\mu^- \to
h^*,H^*,A^*\to \bar{t}c$. The three couplings $h^0\mu^+\mu^-$, $H^0\mu^+\mu^-$
and $A^0\mu^+\mu^-$ behave like $m_{\mu}\tan\beta$ at large
$\tan\beta$ and this might be a source of enhancement in the large
$\tan\beta$ limit. Such enhancement has been already observed
for $\mu^+\mu^-\to W^\pm H^\mp$ and $\mu^+\mu^-\to ZA^0$ \cite{WH}.

In the SM, it is expected that the cross section $\mu^+\mu^- \to \bar{t}c$
is very suppressed as in the case for $e^+e^- \to \bar{t}c$. In the case 
of $\mu^+\mu^- \to \bar{t}c$, the only difference with respect to $e^+e^-
\to \bar{t}c$ is the contribution of Higgs boson s-channel 
$\mu^+\mu^- \to H^*\to
\bar{t}c$. However, in SM the couplings $H^0\mu^+\mu^-$ as well as 
$H^0\bar{t}c$ does not have any enhancement factor like $\tan\beta$.
Therefore, the cross section for
$\mu^+\mu^- \to \bar{t}c$ will be similar in magnitude to the 
$e^+e^- \to \bar{t}c$ cross section.

We show in Fig.~(\ref{muu}) cross sections for $\mu^+\mu^-\to \bar{t}c$
as a function of center of mass energy (left) and as a function of
$\tan\beta$ (right). As expected, the cross sections are enhanced for
large $\tan\beta$ where both $\{h^0,H^0,A^0\}\mu^+\mu^-$ and 
$\{h^0,H^0,A^0\}\bar{t}c$ couplings are enhanced. As  can be seen from
right plot of Fig.~\ref{muu}, the cross section can reach 
a few fb near threshold region for $\tan\beta=65$ and light charged Higgs.
One can see also several kinks, the first one on the left is 
located around $\sqrt{s}\approx M_H=278$ GeV
and is due to s-channel resonance of the 
heavy CP-even Higgs $H^0$. The second kink, located at several 
points where $\sqrt{s}\approx 2M_{H\pm}$, 
is due to the threshold production of pair of charged Higgs.
For Fig.~\ref{muu}, instead of using unitarity constraints 
we have used the  perturbativity constraints on the $\lambda_i$: 
$\lambda_i \la 8 \pi$.
That has been done in order to allow a larger 
reach for $\tan\beta$. For instance,
for charged Higgs of 150 GeV, unitarity constraints are satisfied
only for $\tan\beta\la 48$ while perturbativity constraints on the
$\lambda_i$ are satisfied even for $\tan\beta\approx 100$.
As it can be seen from those plots, the cross section can be larger
than 0.01 fb only for low center of mass energy $\sqrt{s}\la 500$ GeV 
and rather light charged Higgs. For high center of mass energy
$\sqrt{s}\ga 500$ GeV, the cross sections of the s-channel process 
$\mu^+\mu^-\to \bar{t}c$ are reduced.\\
We note that in 2HDM-I, $\mu^+\mu^-\to \bar{t}c$ would not register
any observable signal. This is due to the fact that 
some fermionic couplings  are 
proportional to $\cot^2\beta$, and so 
unacceptably small values of $\tan\beta$
would be required in order to allow observable cross--sections.

\section{Conclusions}
In the framework of 2HDM with natural flavor conservation,
we have studied various top and Higgs flavor 
changing neutral couplings: $t\to c \gamma, cZ, cg, ch^0$  and $\Phi\to
\bar{t}c$. The study has been carried out
taking into account the experimental constraint on the $\rho$ parameter, 
 unitarity constraints, and vacuum stability conditions
 on all scalar quartic couplings $\lambda_i$.
Numerical results for the branching ratios have been presented.
We emphasize the effect coming from bottom Yukawa couplings
and pure trilinear scalar couplings such as $h^0H^+H^-$ and $H^0H^+H^-$.
\\
We have shown that, in 2HDM-II, the branching ratios of 
top FCNC decays $t\to c\gamma , cZ, cg, ch^0 $ as
well as Higgs FCNC $\{h^0 ,H^0\}\to \bar{t}c$ can be orders of
magnitude larger than their SM values. 
Such enhancement requires large $\tan\beta$, rather light charged 
Higgs and not too heavy  CP-even Higgs masses. 
We have shown that $t\to ch^0$ as well as $\{h^0,H^0\}\to \bar{t}c$ 
can reach observable rates at LHC with Branching ratios of the order 
$10^{-5}$ while all the unitarity and vacuum stability 
constraints are fulfilled.  With Branching ratio of this size, 
the production and subsequent decay of the neutral CP-even Higgs 
mediated by FCNC can lead to a few hundred events at LHC with the High 
luminosity option \cite{2hdm4}.
\\
We have also studied the top-charm associated production 
at $e^+e^-$ colliders, its $\gamma\gamma$ option and also at muon colliders.
It has been shown that, in 2HDM-II, 
the cross section of $\gamma\gamma\to \bar{t}c$ 
can be of the order $0.01\to 0.1$ fb near the threshold region
while the corresponding cross sections at $e^+e^-$ 
remain well below $10^{-2}$ fb. 
At muon colliders, the situation is 
slightly better due to Higgs boson $h^0$, $H^0$ and $A^0$ 
s-channel contribution. 
We showed that $\mu^+\mu^-\to \bar{t}c$
production offers an attractive way of 
searching for FCNC at such colliders.
The cross section grows with increasing $\tan\beta$
with values of few fb being attainable for $\tan\beta\ge 48$.

\vspace{1cm}
\noindent
{\Large \bf Acknowledgments}
This work is supported by the Physics Division of
National Center for Theoretical Sciences under a grant from the
National Science Council of Taiwan. We are grateful to A. Akeroyd for
useful discussions and for reading the paper. We thank K. Cheung and
O. Kong for comments about $(g-2)_\mu$. We would like to thank
J. Guasch and S. B\'ejar for exchange of 
informations about Ref.~\cite{2hdm4}.

\end{document}

%% file: hsbb.tex
\unitlength=1bp%

\begin{feynartspicture}(432,504)(6,6.3)

\FADiagram{d$_1$}
\FAProp(0.,10.)(6.5,10.)(0.,){/ScalarDash}{0}
\FALabel(3.25,9.18)[t]{$\Phi$}
\FAProp(20.,15.)(13.,14.)(0.,){/Straight}{-1}
\FALabel(16.2808,15.5544)[b]{$f_1$}
\FAProp(20.,5.)(13.,6.)(0.,){/Straight}{1}
\FALabel(16.2808,4.44558)[t]{$f_2$}
\FAProp(6.5,10.)(13.,14.)(0.,){/Straight}{1}
\FALabel(9.20801,13.1807)[br]{$f^\prime_i$}
\FAProp(6.5,10.)(13.,6.)(0.,){/Straight}{-1}
\FALabel(9.20801,6.81927)[tr]{$f^\prime_i$}
\FAProp(13.,14.)(13.,6.)(0.,){/ScalarDash}{1}
\FALabel(14.274,10.)[l]{$G$}
\FAVert(6.5,10.){0}
\FAVert(13.,14.){0}
\FAVert(13.,6.){0}

\FADiagram{d$_2$}
\FAProp(0.,10.)(6.5,10.)(0.,){/ScalarDash}{0}
\FALabel(3.25,9.18)[t]{$\Phi$}
\FAProp(20.,15.)(13.,14.)(0.,){/Straight}{-1}
\FALabel(16.2808,15.5544)[b]{$f_1$}
\FAProp(20.,5.)(13.,6.)(0.,){/Straight}{1}
\FALabel(16.2808,4.44558)[t]{$f_2$}
\FAProp(6.5,10.)(13.,14.)(0.,){/ScalarDash}{-1}
\FALabel(9.20801,13.1807)[br]{$G$}
\FAProp(6.5,10.)(13.,6.)(0.,){/ScalarDash}{1}
\FALabel(9.20801,6.81927)[tr]{$G$}
\FAProp(13.,14.)(13.,6.)(0.,){/Straight}{-1}
\FALabel(14.274,10.)[l]{$f^\prime_i$}
\FAVert(6.5,10.){0}
\FAVert(13.,14.){0}
\FAVert(13.,6.){0}

\FADiagram{d$_3$}
\FAProp(0.,10.)(6.5,10.)(0.,){/ScalarDash}{0}
\FALabel(3.25,9.18)[t]{$\Phi$}
\FAProp(20.,15.)(13.,14.)(0.,){/Straight}{-1}
\FALabel(16.2808,15.5544)[b]{$f_1$}
\FAProp(20.,5.)(13.,6.)(0.,){/Straight}{1}
\FALabel(16.2808,4.44558)[t]{$f_2$}
\FAProp(6.5,10.)(13.,14.)(0.,){/Straight}{1}
\FALabel(9.20801,13.1807)[br]{$f^\prime_i$}
\FAProp(6.5,10.)(13.,6.)(0.,){/Straight}{-1}
\FALabel(9.20801,6.81927)[tr]{$f^\prime_i$}
\FAProp(13.,14.)(13.,6.)(0.,){/Sine}{1}
\FALabel(14.274,10.)[l]{$W$}
\FAVert(6.5,10.){0}
\FAVert(13.,14.){0}
\FAVert(13.,6.){0}

\FADiagram{d$_4$}
\FAProp(0.,10.)(6.5,10.)(0.,){/ScalarDash}{0}
\FALabel(3.25,9.18)[t]{$\Phi$}
\FAProp(20.,15.)(13.,14.)(0.,){/Straight}{-1}
\FALabel(16.2808,15.5544)[b]{$f_1$}
\FAProp(20.,5.)(13.,6.)(0.,){/Straight}{1}
\FALabel(16.2808,4.44558)[t]{$f_2$}
\FAProp(6.5,10.)(13.,14.)(0.,){/ScalarDash}{-1}
\FALabel(9.20801,13.1807)[br]{$G$}
\FAProp(6.5,10.)(13.,6.)(0.,){/Sine}{1}
\FALabel(9.20801,6.81927)[tr]{$W$}
\FAProp(13.,14.)(13.,6.)(0.,){/Straight}{-1}
\FALabel(14.274,10.)[l]{$f^\prime_i$}
\FAVert(6.5,10.){0}
\FAVert(13.,14.){0}
\FAVert(13.,6.){0}

\FADiagram{d$_5$}
\FAProp(0.,10.)(6.5,10.)(0.,){/ScalarDash}{0}
\FALabel(3.25,9.18)[t]{$\Phi$}
\FAProp(20.,15.)(13.,14.)(0.,){/Straight}{-1}
\FALabel(16.2808,15.5544)[b]{$f_1$}
\FAProp(20.,5.)(13.,6.)(0.,){/Straight}{1}
\FALabel(16.2808,4.44558)[t]{$f_2$}
\FAProp(6.5,10.)(13.,14.)(0.,){/Sine}{-1}
\FALabel(9.20801,13.1807)[br]{$W$}
\FAProp(6.5,10.)(13.,6.)(0.,){/ScalarDash}{1}
\FALabel(9.20801,6.81927)[tr]{$G$}
\FAProp(13.,14.)(13.,6.)(0.,){/Straight}{-1}
\FALabel(14.274,10.)[l]{$f^\prime_i$}
\FAVert(6.5,10.){0}
\FAVert(13.,14.){0}
\FAVert(13.,6.){0}

\FADiagram{d$_6$}
\FAProp(0.,10.)(6.5,10.)(0.,){/ScalarDash}{0}
\FALabel(3.25,9.18)[t]{$\Phi$}
\FAProp(20.,15.)(13.,14.)(0.,){/Straight}{-1}
\FALabel(16.2808,15.5544)[b]{$f_1$}
\FAProp(20.,5.)(13.,6.)(0.,){/Straight}{1}
\FALabel(16.2808,4.44558)[t]{$f_2$}
\FAProp(6.5,10.)(13.,14.)(0.,){/Sine}{-1}
\FALabel(9.20801,13.1807)[br]{$W$}
\FAProp(6.5,10.)(13.,6.)(0.,){/Sine}{1}
\FALabel(9.20801,6.81927)[tr]{$W$}
\FAProp(13.,14.)(13.,6.)(0.,){/Straight}{-1}
\FALabel(14.274,10.)[l]{$f^\prime_i$}
\FAVert(6.5,10.){0}
\FAVert(13.,14.){0}
\FAVert(13.,6.){0}

\FADiagram{d$_7$}
\FAProp(0.,10.)(11.,10.)(0.,){/ScalarDash}{0}
\FALabel(5.5,9.18)[t]{$\Phi$}
\FAProp(20.,15.)(11.,10.)(0.,){/Straight}{-1}
\FALabel(15.2273,13.3749)[br]{$f_1$}
\FAProp(20.,5.)(17.3,6.5)(0.,){/Straight}{1}
\FALabel(18.9227,6.62494)[bl]{$f_2$}
\FAProp(11.,10.)(13.7,8.5)(0.,){/Straight}{-1}
\FALabel(12.0773,8.37506)[tr]{$f_1$}
\FAProp(17.3,6.5)(13.7,8.5)(-0.8,){/Straight}{1}
\FALabel(14.4273,5.18506)[tr]{$f^\prime_i$}
\FAProp(17.3,6.5)(13.7,8.5)(0.8,){/ScalarDash}{-1}
\FALabel(16.5727,9.81494)[bl]{$G$}
\FAVert(11.,10.){0}
\FAVert(17.3,6.5){0}
\FAVert(13.7,8.5){0}

\FADiagram{d$_{8}$}
\FAProp(0.,10.)(11.,10.)(0.,){/ScalarDash}{0}
\FALabel(5.5,9.18)[t]{$\Phi$}
\FAProp(20.,15.)(11.,10.)(0.,){/Straight}{-1}
\FALabel(15.2273,13.3749)[br]{$f_1$}
\FAProp(20.,5.)(17.3,6.5)(0.,){/Straight}{1}
\FALabel(18.9227,6.62494)[bl]{$f_2$}
\FAProp(11.,10.)(13.7,8.5)(0.,){/Straight}{-1}
\FALabel(12.0773,8.37506)[tr]{$f_1$}
\FAProp(17.3,6.5)(13.7,8.5)(-0.8,){/Straight}{1}
\FALabel(14.4273,5.18506)[tr]{$f^\prime_i$}
\FAProp(17.3,6.5)(13.7,8.5)(0.8,){/Sine}{-1}
\FALabel(16.5727,9.81494)[bl]{$W$}
\FAVert(11.,10.){0}
\FAVert(17.3,6.5){0}
\FAVert(13.7,8.5){0}

\FADiagram{d$_{9}$}
\FAProp(0.,10.)(11.,10.)(0.,){/ScalarDash}{0}
\FALabel(5.5,9.18)[t]{$\Phi$}
\FAProp(20.,15.)(17.3,13.5)(0.,){/Straight}{-1}
\FALabel(18.3773,15.1249)[br]{$f_1$}
\FAProp(20.,5.)(11.,10.)(0.,){/Straight}{1}
\FALabel(15.2273,6.62506)[tr]{$f_2$}
\FAProp(11.,10.)(13.7,11.5)(0.,){/Straight}{1}
\FALabel(12.0773,11.6249)[br]{$f_2$}
\FAProp(17.3,13.5)(13.7,11.5)(-0.8,){/Straight}{-1}
\FALabel(16.5727,10.1851)[tl]{$f^\prime_i$}
\FAProp(17.3,13.5)(13.7,11.5)(0.8,){/ScalarDash}{1}
\FALabel(14.4273,14.8149)[br]{$G$}
\FAVert(11.,10.){0}
\FAVert(17.3,13.5){0}
\FAVert(13.7,11.5){0}

\FADiagram{d$_{10}$}
\FAProp(0.,10.)(11.,10.)(0.,){/ScalarDash}{0}
\FALabel(5.5,9.18)[t]{$\Phi$}
\FAProp(20.,15.)(17.3,13.5)(0.,){/Straight}{-1}
\FALabel(18.3773,15.1249)[br]{$f_1$}
\FAProp(20.,5.)(11.,10.)(0.,){/Straight}{1}
\FALabel(15.2273,6.62506)[tr]{$f_2$}
\FAProp(11.,10.)(13.7,11.5)(0.,){/Straight}{1}
\FALabel(12.0773,11.6249)[br]{$f_2$}
\FAProp(17.3,13.5)(13.7,11.5)(-0.8,){/Straight}{-1}
\FALabel(16.5727,10.1851)[tl]{$f^\prime_i$}
\FAProp(17.3,13.5)(13.7,11.5)(0.8,){/Sine}{1}
\FALabel(14.4273,14.8149)[br]{$W$}
\FAVert(11.,10.){0}
\FAVert(17.3,13.5){0}
\FAVert(13.7,11.5){0}

\FADiagram{d$_{11}$}
\FAProp(0.,10.)(6.5,10.)(0.,){/ScalarDash}{0}
\FALabel(3.25,9.18)[t]{$\Phi$}
\FAProp(20.,15.)(13.,14.)(0.,){/Straight}{-1}
\FALabel(16.2808,15.5544)[b]{$f_1$}
\FAProp(20.,5.)(13.,6.)(0.,){/Straight}{1}
\FALabel(16.2808,4.44558)[t]{$f_2$}
\FAProp(6.5,10.)(13.,14.)(0.,){/Straight}{1}
\FALabel(9.20801,13.1807)[br]{$f^\prime_i$}
\FAProp(6.5,10.)(13.,6.)(0.,){/Straight}{-1}
\FALabel(9.20801,6.81927)[tr]{$f^\prime_i$}
\FAProp(13.,14.)(13.,6.)(0.,){/ScalarDash}{1}
\FALabel(14.274,10.)[l]{$H$}
\FAVert(6.5,10.){0}
\FAVert(13.,14.){0}
\FAVert(13.,6.){0}

\FADiagram{d$_{12}$}
\FAProp(0.,10.)(6.5,10.)(0.,){/ScalarDash}{0}
\FALabel(3.25,9.18)[t]{$\Phi$}
\FAProp(20.,15.)(13.,14.)(0.,){/Straight}{-1}
\FALabel(16.2808,15.5544)[b]{$f_1$}
\FAProp(20.,5.)(13.,6.)(0.,){/Straight}{1}
\FALabel(16.2808,4.44558)[t]{$f_2$}
\FAProp(6.5,10.)(13.,14.)(0.,){/ScalarDash}{-1}
\FALabel(9.20801,13.1807)[br]{$H$}
\FAProp(6.5,10.)(13.,6.)(0.,){/ScalarDash}{1}
\FALabel(9.20801,6.81927)[tr]{$H$}
\FAProp(13.,14.)(13.,6.)(0.,){/Straight}{-1}
\FALabel(14.274,10.)[l]{$f^\prime_i$}
\FAVert(6.5,10.){0}
\FAVert(13.,14.){0}
\FAVert(13.,6.){0}

\FADiagram{d$_{13}$}
\FAProp(0.,10.)(6.5,10.)(0.,){/ScalarDash}{0}
\FALabel(3.25,9.18)[t]{$\Phi$}
\FAProp(20.,15.)(13.,14.)(0.,){/Straight}{-1}
\FALabel(16.2808,15.5544)[b]{$f_1$}
\FAProp(20.,5.)(13.,6.)(0.,){/Straight}{1}
\FALabel(16.2808,4.44558)[t]{$f_2$}
\FAProp(6.5,10.)(13.,14.)(0.,){/ScalarDash}{-1}
\FALabel(9.20801,13.1807)[br]{$H$}
\FAProp(6.5,10.)(13.,6.)(0.,){/ScalarDash}{1}
\FALabel(9.20801,6.81927)[tr]{$G$}
\FAProp(13.,14.)(13.,6.)(0.,){/Straight}{-1}
\FALabel(14.274,10.)[l]{$f^\prime_i$}
\FAVert(6.5,10.){0}
\FAVert(13.,14.){0}
\FAVert(13.,6.){0}

\FADiagram{d$_{14}$}
\FAProp(0.,10.)(6.5,10.)(0.,){/ScalarDash}{0}
\FALabel(3.25,9.18)[t]{$\Phi$}
\FAProp(20.,15.)(13.,14.)(0.,){/Straight}{-1}
\FALabel(16.2808,15.5544)[b]{$f_1$}
\FAProp(20.,5.)(13.,6.)(0.,){/Straight}{1}
\FALabel(16.2808,4.44558)[t]{$f_2$}
\FAProp(6.5,10.)(13.,14.)(0.,){/ScalarDash}{-1}
\FALabel(9.20801,13.1807)[br]{$G$}
\FAProp(6.5,10.)(13.,6.)(0.,){/ScalarDash}{1}
\FALabel(9.20801,6.81927)[tr]{$H$}
\FAProp(13.,14.)(13.,6.)(0.,){/Straight}{-1}
\FALabel(14.274,10.)[l]{$f^\prime_i$}
\FAVert(6.5,10.){0}
\FAVert(13.,14.){0}
\FAVert(13.,6.){0}

\FADiagram{d$_{15}$}
\FAProp(0.,10.)(6.5,10.)(0.,){/ScalarDash}{0}
\FALabel(3.25,9.18)[t]{$\Phi$}
\FAProp(20.,15.)(13.,14.)(0.,){/Straight}{-1}
\FALabel(16.2808,15.5544)[b]{$f_1$}
\FAProp(20.,5.)(13.,6.)(0.,){/Straight}{1}
\FALabel(16.2808,4.44558)[t]{$f_2$}
\FAProp(6.5,10.)(13.,14.)(0.,){/ScalarDash}{-1}
\FALabel(9.20801,13.1807)[br]{$H$}
\FAProp(6.5,10.)(13.,6.)(0.,){/Sine}{1}
\FALabel(9.20801,6.81927)[tr]{$W$}
\FAProp(13.,14.)(13.,6.)(0.,){/Straight}{-1}
\FALabel(14.274,10.)[l]{$f^\prime_i$}
\FAVert(6.5,10.){0}
\FAVert(13.,14.){0}
\FAVert(13.,6.){0}

\FADiagram{d$_{16}$}
\FAProp(0.,10.)(6.5,10.)(0.,){/ScalarDash}{0}
\FALabel(3.25,9.18)[t]{$\Phi$}
\FAProp(20.,15.)(13.,14.)(0.,){/Straight}{-1}
\FALabel(16.2808,15.5544)[b]{$f_1$}
\FAProp(20.,5.)(13.,6.)(0.,){/Straight}{1}
\FALabel(16.2808,4.44558)[t]{$f_2$}
\FAProp(6.5,10.)(13.,14.)(0.,){/Sine}{-1}
\FALabel(9.20801,13.1807)[br]{$W$}
\FAProp(6.5,10.)(13.,6.)(0.,){/ScalarDash}{1}
\FALabel(9.20801,6.81927)[tr]{$H$}
\FAProp(13.,14.)(13.,6.)(0.,){/Straight}{-1}
\FALabel(14.274,10.)[l]{$f^\prime_i$}
\FAVert(6.5,10.){0}
\FAVert(13.,14.){0}
\FAVert(13.,6.){0}

\FADiagram{d$_{17}$}
\FAProp(0.,10.)(11.,10.)(0.,){/ScalarDash}{0}
\FALabel(5.5,9.18)[t]{$\Phi$}
\FAProp(20.,15.)(11.,10.)(0.,){/Straight}{-1}
\FALabel(15.2273,13.3749)[br]{$f_1$}
\FAProp(20.,5.)(17.3,6.5)(0.,){/Straight}{1}
\FALabel(18.9227,6.62494)[bl]{$f_2$}
\FAProp(11.,10.)(13.7,8.5)(0.,){/Straight}{-1}
\FALabel(12.0773,8.37506)[tr]{$f_1$}
\FAProp(17.3,6.5)(13.7,8.5)(-0.8,){/Straight}{1}
\FALabel(14.4273,5.18506)[tr]{$f^\prime_i$}
\FAProp(17.3,6.5)(13.7,8.5)(0.8,){/ScalarDash}{-1}
\FALabel(16.5727,9.81494)[bl]{$H$}
\FAVert(11.,10.){0}
\FAVert(17.3,6.5){0}
\FAVert(13.7,8.5){0}

\FADiagram{d$_{18}$}
\FAProp(0.,10.)(11.,10.)(0.,){/ScalarDash}{0}
\FALabel(5.5,9.18)[t]{$\Phi$}
\FAProp(20.,15.)(17.3,13.5)(0.,){/Straight}{-1}
\FALabel(18.3773,15.1249)[br]{$f_1$}
\FAProp(20.,5.)(11.,10.)(0.,){/Straight}{1}
\FALabel(15.2273,6.62506)[tr]{$f_2$}
\FAProp(11.,10.)(13.7,11.5)(0.,){/Straight}{1}
\FALabel(12.0773,11.6249)[br]{$f_2$}
\FAProp(17.3,13.5)(13.7,11.5)(-0.8,){/Straight}{-1}
\FALabel(16.5727,10.1851)[tl]{$f^\prime_i$}
\FAProp(17.3,13.5)(13.7,11.5)(0.8,){/ScalarDash}{1}
\FALabel(14.4273,14.8149)[br]{$H$}
\FAVert(11.,10.){0}
\FAVert(17.3,13.5){0}
\FAVert(13.7,11.5){0}

\FADiagram{}

\FADiagram{}

\FADiagram{}

\FADiagram{}

\FADiagram{}

\FADiagram{}

\FADiagram{}

\FADiagram{}

\FADiagram{}

\FADiagram{}

\FADiagram{}

\FADiagram{}

\FADiagram{}

\FADiagram{}
\end{feynartspicture}

%% file: ee.tex
\unitlength=1bp%

\begin{feynartspicture}(432,504)(5,5.3)
\FADiagram{$d_1$}
\FAProp(0.,15.)(4.,10.)(0.,){/Straight}{0}
\FAProp(0.,5.)(4.,10.)(0.,){/Straight}{0}
\FAProp(20.,15.)(16.,13.5)(0.,){/Straight}{0}
\FAProp(20.,5.)(16.,6.5)(0.,){/Straight}{0}
\FAProp(4.,10.)(10.,10.)(0.,){/Straight}{0}
\FAProp(16.,13.5)(16.,6.5)(0.,){/Straight}{0}
\FAProp(16.,13.5)(10.,10.)(0.,){/Straight}{0}
\FAProp(16.,6.5)(10.,10.)(0.,){/Straight}{0}
\FAVert(4.,10.){0}
\FAVert(16.,13.5){0}
\FAVert(16.,6.5){0}
\FAVert(10.,10.){0}

\FADiagram{$d_2$}
\FAProp(0.,15.)(6.,10.)(0.,){/Straight}{0}
\FAProp(0.,5.)(6.,10.)(0.,){/Straight}{0}
\FAProp(20.,15.)(14.,10.)(0.,){/Straight}{0}
\FAProp(20.,5.)(18.2,6.5)(0.,){/Straight}{0}
\FAProp(6.,10.)(14.,10.)(0.,){/Straight}{0}
\FAProp(14.,10.)(15.8,8.5)(0.,){/Straight}{0}
\FAProp(18.2,6.5)(15.8,8.5)(-0.8,){/Straight}{0}
\FAProp(18.2,6.5)(15.8,8.5)(0.8,){/Straight}{0}
\FAVert(6.,10.){0}
\FAVert(14.,10.){0}
\FAVert(18.2,6.5){0}
\FAVert(15.8,8.5){0}

\FADiagram{$d_3$}
\FAProp(0.,15.)(6.,10.)(0.,){/Straight}{0}
\FAProp(0.,5.)(6.,10.)(0.,){/Straight}{0}
\FAProp(20.,15.)(18.2,13.5)(0.,){/Straight}{0}
\FAProp(20.,5.)(14.,10.)(0.,){/Straight}{0}
\FAProp(6.,10.)(14.,10.)(0.,){/Straight}{0}
\FAProp(14.,10.)(15.8,11.5)(0.,){/Straight}{0}
\FAProp(18.2,13.5)(15.8,11.5)(-0.8,){/Straight}{0}
\FAProp(18.2,13.5)(15.8,11.5)(0.8,){/Straight}{0}
\FAVert(6.,10.){0}
\FAVert(18.2,13.5){0}
\FAVert(14.,10.){0}
\FAVert(15.8,11.5){0}

\FADiagram{$d_4$}
\FAProp(0.,15.)(6.5,13.5)(0.,){/Straight}{0}
\FAProp(0.,5.)(6.5,6.5)(0.,){/Straight}{0}
\FAProp(20.,15.)(13.5,13.5)(0.,){/Straight}{0}
\FAProp(20.,5.)(13.5,6.5)(0.,){/Straight}{0}
\FAProp(6.5,13.5)(6.5,6.5)(0.,){/Straight}{0}
\FAProp(6.5,13.5)(13.5,13.5)(0.,){/Straight}{0}
\FAProp(6.5,6.5)(13.5,6.5)(0.,){/Straight}{0}
\FAProp(13.5,13.5)(13.5,6.5)(0.,){/Straight}{0}
\FAVert(6.5,13.5){0}
\FAVert(6.5,6.5){0}
\FAVert(13.5,13.5){0}
\FAVert(13.5,6.5){0}
\FALabel(38.5,11.5)[tr]{+ crossed}

\end{feynartspicture}

%% file: gg.tex
\unitlength=1bp%

\begin{feynartspicture}(432,504)(5,5.3)

\FADiagram{$d_1$}
\FAProp(0.,15.)(8.,10.)(0.,){/Straight}{0}
\FAProp(0.,5.)(8.,10.)(0.,){/Straight}{0}
\FAProp(20.,15.)(14.,13.5)(0.,){/Straight}{0}
\FAProp(20.,5.)(14.,6.5)(0.,){/Straight}{0}
\FAProp(14.,13.5)(14.,6.5)(0.,){/Straight}{0}
\FAProp(14.,13.5)(8.,10.)(0.,){/Straight}{0}
\FAProp(14.,6.5)(8.,10.)(0.,){/Straight}{0}
\FAVert(14.,13.5){0}
\FAVert(14.,6.5){0}
\FAVert(8.,10.){0}

\FADiagram{$d_2$}
\FAProp(0.,15.)(10.,14.5)(0.,){/Straight}{0}
\FAProp(0.,5.)(6.5,5.5)(0.,){/Straight}{0}
\FAProp(20.,15.)(10.,14.5)(0.,){/Straight}{0}
\FAProp(20.,5.)(13.5,5.5)(0.,){/Straight}{0}
\FAProp(10.,14.5)(10.,11.)(0.,){/Straight}{0}
\FAProp(6.5,5.5)(13.5,5.5)(0.,){/Straight}{0}
\FAProp(6.5,5.5)(10.,11.)(0.,){/Straight}{0}
\FAProp(13.5,5.5)(10.,11.)(0.,){/Straight}{0}
\FAVert(10.,14.5){0}
\FAVert(6.5,5.5){0}
\FAVert(13.5,5.5){0}
\FAVert(10.,11.){0}

\FADiagram{$d_3$}
\FAProp(0.,15.)(6.5,14.5)(0.,){/Straight}{0}
\FAProp(0.,5.)(10.,5.5)(0.,){/Straight}{0}
\FAProp(20.,15.)(13.5,14.5)(0.,){/Straight}{0}
\FAProp(20.,5.)(10.,5.5)(0.,){/Straight}{0}
\FAProp(10.,5.5)(10.,8.5)(0.,){/Straight}{0}
\FAProp(6.5,14.5)(13.5,14.5)(0.,){/Straight}{0}
\FAProp(6.5,14.5)(10.,8.5)(0.,){/Straight}{0}
\FAProp(13.5,14.5)(10.,8.5)(0.,){/Straight}{0}
\FAVert(6.5,14.5){0}
\FAVert(10.,5.5){0}
\FAVert(13.5,14.5){0}
\FAVert(10.,8.5){0}

\FADiagram{$d_4$}
\FAProp(0.,15.)(6.5,13.5)(0.,){/Straight}{0}
\FAProp(0.,5.)(6.5,6.5)(0.,){/Straight}{0}
\FAProp(20.,15.)(13.5,13.5)(0.,){/Straight}{0}
\FAProp(20.,5.)(13.5,6.5)(0.,){/Straight}{0}
\FAProp(6.5,13.5)(6.5,6.5)(0.,){/Straight}{0}
\FAProp(6.5,13.5)(13.5,13.5)(0.,){/Straight}{0}
\FAProp(6.5,6.5)(13.5,6.5)(0.,){/Straight}{0}
\FAProp(13.5,13.5)(13.5,6.5)(0.,){/Straight}{0}
\FAVert(6.5,13.5){0}
\FAVert(6.5,6.5){0}
\FAVert(13.5,13.5){0}
\FAVert(13.5,6.5){0}

\FADiagram{$d_5$}
\FAProp(0.,15.)(13.5,13.)(0.,){/Straight}{0}
\FAProp(0.,5.)(6.5,6.)(0.,){/Straight}{0}
\FAProp(20.,15.)(6.5,13.)(0.,){/Straight}{0}
\FAProp(20.,5.)(13.5,6.)(0.,){/Straight}{0}
\FAProp(13.5,13.)(6.5,13.)(0.,){/Straight}{0}
\FAProp(13.5,13.)(13.5,6.)(0.,){/Straight}{0}
\FAProp(6.5,6.)(6.5,13.)(0.,){/Straight}{0}
\FAProp(6.5,6.)(13.5,6.)(0.,){/Straight}{0}
\FAVert(13.5,13.){0}
\FAVert(6.5,6.){0}
\FAVert(6.5,13.){0}
\FAVert(13.5,6.){0}

\FADiagram{$d_6$}
\FAProp(0.,15.)(10.,14.5)(0.,){/Straight}{0}
\FAProp(0.,5.)(10.,5.5)(0.,){/Straight}{0}
\FAProp(20.,15.)(10.,14.5)(0.,){/Straight}{0}
\FAProp(20.,5.)(10.,5.5)(0.,){/Straight}{0}
\FAProp(10.,14.5)(10.,12.)(0.,){/Straight}{0}
\FAProp(10.,5.5)(10.,8.)(0.,){/Straight}{0}
\FAProp(10.,12.)(10.,8.)(1.,){/Straight}{0}
\FAProp(10.,12.)(10.,8.)(-1.,){/Straight}{0}
\FAVert(10.,14.5){0}
\FAVert(10.,5.5){0}
\FAVert(10.,12.){0}
\FAVert(10.,8.){0}

\FADiagram{$d_7$ }
\FAProp(0.,15.)(10.,14.)(0.,){/Straight}{0}
\FAProp(0.,5.)(10.,6.)(0.,){/Straight}{0}
\FAProp(20.,15.)(10.,14.)(0.,){/Straight}{0}
\FAProp(20.,5.)(16.95,5.3)(0.,){/Straight}{0}
\FAProp(10.,14.)(10.,6.)(0.,){/Straight}{0}
\FAProp(10.,6.)(13.05,5.75)(0.,){/Straight}{0}
\FAProp(16.95,5.3)(13.05,5.75)(-0.8,){/Straight}{0}
\FAProp(16.95,5.3)(13.05,5.75)(0.8,){/Straight}{0}
\FAVert(10.,14.){0}
\FAVert(10.,6.){0}
\FAVert(16.95,5.3){0}
\FAVert(13.05,5.75){0}

\FADiagram{$d_8$}
\FAProp(0.,15.)(10.,14.)(0.,){/Straight}{0}
\FAProp(0.,5.)(10.,6.)(0.,){/Straight}{0}
\FAProp(20.,15.)(17.,14.7)(0.,){/Straight}{0}
\FAProp(20.,5.)(10.,6.)(0.,){/Straight}{0}
\FAProp(10.,14.)(10.,6.)(0.,){/Straight}{0}
\FAProp(10.,14.)(13.,14.3)(0.,){/Straight}{0}
\FAProp(17.,14.7)(13.,14.3)(-0.8,){/Straight}{0}
\FAProp(17.,14.7)(13.,14.3)(0.8,){/Straight}{0}
\FAVert(10.,14.){0}
\FAVert(10.,6.){0}
\FAVert(17.,14.7){0}
\FAVert(13.,14.3){0}
\FALabel(38.5,11.5)[tr]{+ crossed}

\end{feynartspicture}